\documentclass[aps,prx,noshowpacs,twocolumn,nobibnotes]{revtex4-1}

\usepackage{amssymb}
\usepackage{graphicx}
\usepackage{amsmath,amsfonts}
\usepackage{hyperref}
\usepackage[utf8] {inputenc}
\usepackage{xcolor}  
\usepackage[normalem]{ulem}
\usepackage[T1]{fontenc} 
\usepackage{multirow}
\usepackage{booktabs}
\usepackage{threeparttable}
\usepackage{tabularx}
\usepackage{subcaption} 
\usepackage{ragged2e} 

\usepackage{listings}
\usepackage{caption}

\definecolor{codebg}{rgb}{0.96,0.96,0.96}
\definecolor{codekw}{rgb}{0.00,0.20,0.60}
\definecolor{codestr}{rgb}{0.58,0.00,0.82}
\definecolor{codecmt}{rgb}{0.65,0.16,0.16}
\definecolor{codecmt}{rgb}{0.72,0.28,0.0}

\definecolor{codenums}{rgb}{0.45,0.45,0.45}

\definecolor{codefunc}{rgb}{0.58,0.00,0.82}  

\lstdefinestyle{mystyle}{
  backgroundcolor=\color{codebg},
  basicstyle=\ttfamily\footnotesize,
  keywordstyle=\bfseries\color{codekw},
  stringstyle=\color{codestr},
  commentstyle=\itshape\color{codecmt},
  numberstyle=\tiny\color{codenums},
  numbers=left,
  numbersep=8pt,
  frame=single,
  frameround=tttt,
  breaklines=true,
  tabsize=4,
  showstringspaces=false,
  columns=fullflexible,
  aboveskip=0pt,
  belowskip=0pt,
  captionpos=b,
  emph={evaluate,equation_v0}, 
  emphstyle=\color{codefunc}
}
\renewcommand{\thetable}{\arabic{table}}
\renewcommand{\thefigure}{\arabic{figure}}

\makeatletter
\renewcommand{\fnum@table}[1]{\textbf{Table~\thetable}:}
\renewcommand{\fnum@figure}[1]{\textbf{Figure~\thefigure}:}
\makeatother


\begin{document}
\title{Foundation models for equation discovery in high energy physics}
\author{Manuel Morales-Alvarado}
\affiliation{INFN, Sezione di Trieste, SISSA,
Via Bonomea 265, 34136, Trieste, Italy}


\begin{abstract}
Foundation models, large machine learning models trained on broad, multimodal datasets, have been gaining increasing attention in scientific applications due to their strong performance on diverse downstream tasks. Large Language Models (LLMs), a prominent instance of foundation models, have achieved remarkable success in tasks such as text and image generation. In this work, we investigate their potential for equation discovery in high energy physics, focusing on symbolic regression. We apply the LLM-SR methodology both to benchmark problems of equation recovery in lepton angular distributions and to the discovery of functional forms for angular coefficients in electroweak boson production at the Large Hadron Collider, observables of high phenomenological relevance for which no closed-form expressions are known from first principles. Our results demonstrate that LLM-SR can uncover compact, accurate, and interpretable equations across in-domain and out-of-domain kinematic regions, effectively incorporating embedded scientific knowledge and offering a promising new approach to equation discovery in high energy physics.
\end{abstract}
\maketitle
\let\oldaddcontentsline\addcontentsline
\renewcommand{\addcontentsline}[3]{}

\section{Introduction}

Foundation models represent a recent paradigm shift in machine learning, characterised by large-scale pretraining on broad, multimodal datasets followed by adaptation to diverse downstream tasks. Rather than being designed for a single application, these models provide flexible representations that can be specialised efficiently through finetuning for downstream tasks. Large language models (LLMs)~\cite{bommasani2021opportunities,wei2022emergent} are a particularly successful example of foundation models, having demonstrated excellent capabilities in tasks such as text and image generation, and increasingly in scientific problem solving~\cite{yamada2025aiscientistv2workshoplevelautomated}. 

In high energy physics (HEP), foundation models have recently emerged as powerful tools with applications spanning jet tagging, event classification, accelerator tuning, and self-supervised representation learning~\cite{Heneka:2025fpe,Kaiser:2024lkg,Birk:2024knn,Qu:2022mxj,Golling:2024abg,Harris:2024sra,Mikuni:2024qsr,Ho:2024qyf,Bardhan:2025icr,Hallin:2025ywf}. This broad activity illustrates the increasing relevance of foundation models in HEP and motivates their application to tasks that go beyond standard classification or generation.

In this paper, we investigate the use of foundation models for \emph{equation discovery} in HEP. The problem is naturally framed in terms of symbolic regression (SR), a machine learning technique that aims to uncover explicit analytical expressions from data without fixing the functional form~\cite{petersen2019deep,mundhenk2021symbolic,Udrescu:2019mnk,pysr}. SR plays an important role in scientific discovery by providing compact and interpretable models that bridge the gap between black-box methods and physical understanding. In HEP, such expressions provide flexible alternatives to fixed-form functions used to model signal and background, enable smooth interpolation of binned observables, and yield analytical approximations where closed-form solutions are unknown. Applications include Higgs-sensitive kinematic variables, optimal observables, analytic simplifications, and more recent studies in collider phenomenology, BSM searches, and fast inference~\cite{Choi:2010wa,Butter:2021rvz,Dersy:2022bym,Dong:2022trn,Alnuqaydan:2022ncd,AbdusSalam:2024obf,Tsoi:2023isc,Tsoi:2024pbn,Soybelman:2024mbv,Dotson:2025omi,Bahl:2025jtk,Vent:2025ddm,Morales-Alvarado:2024jrk,Bendavid:2025urn}. The field has advanced considerably thanks to new tools, e.g.~\cite{pysr,Tsoi:2024ypg}, which make the search for symbolic models significantly more tractable despite the underlying computational challenges~\cite{virgolin2022symbolicregressionnphard}.

In this project, we employ the LLM-SR methodology~\cite{llmsr}, with GPT-4o as the backbone model. Unlike traditional approaches, LLM-SR does not rely on a predefined operator basis but leverages the extensive knowledge embedded in the backbone LLM, providing greater flexibility in exploring the space of functions. The framework treats equations as Python~\cite{python} programs of the form \texttt{def f(x, params): ... return y}, which the LLM proposes as skeleton programs with placeholder parameters, to be optimised over a fixed number of iterations (which in this study we take to be 100 in all cases) and subject to a maximum number of free parameters, both defined by the user. At each iteration, a batch of such skeletons is generated, their parameters are numerically optimised using standard libraries such as \texttt{numpy/scipy}~\cite{numpy,scipy} or \texttt{torch}~\cite{pytorch}, and the best-performing hypotheses are stored in an evolving experience buffer. This iterative process refines both the equation structures and their coefficients by combining domain knowledge encoded through informative prompts in the program, comments in the code, and prior equations, with data-driven feedback. For the detailed description of the algorithm, including the hypothesis generation, optimisation, and experience management steps, we refer the reader to the original reference above.

To benchmark the approach, we first apply LLM-SR to equation \textit{recovery} in lepton angular distributions, where the underlying physical law is known. We then move to equation \textit{discovery} for angular coefficients in electroweak boson production at the Large Hadron Collider (LHC), quantities of high phenomenological relevance for which no closed-form analytical expressions exist. These case studies allow us to assess both the interpretability and the accuracy of the discovered equations, and to compare the methodology against other state-of-the-art SR results.

The remainder of this paper is structured as follows. In Section~\ref{sec:recovery}, we present the recovery of known equations in lepton angular distributions as a benchmark. 
In Section~\ref{sec:discovery}, we apply the method to angular coefficients in one, two, and three kinematic dimensions, examining the impact of the number of free parameters and the role of domain-informed program priors. Section~\ref{sec:conclusions} presents our conclusions, and the Appendix provides example programs for equation discovery.

\section{Equation recovery}
\label{sec:recovery}

To validate the LLM-SR methodology in a high-energy physics setting, it is instructive to test whether it can recover known equations from simulated datasets that are subject to statistical fluctuations. As a benchmark, we generate pseudodata for $e^+ e^- \to \mu^+ \mu^-$ at $\sqrt{s} = 1$~TeV using \textsc{MadGraph5\_aMC@NLO}~\cite{Alwall:2014hca,Frederix:2018nkq}, without applying cuts on the transverse momenta, rapidities, or separation distances of the outgoing leptons. The process provides a setting in which the leading-order QED prediction for the angular distribution is known  
\begin{equation}
    \label{eq:lepton}
    \frac{d\sigma}{d\Omega} = \frac{\alpha^2}{4s}\left(1 + \cos^2\theta\right),
\end{equation}
where $\theta$ is the scattering angle between the incoming electron and the outgoing muon, $d\Omega = d(\cos \theta)\, d\phi$ is the differential solid angle, and $\alpha$ is the QED coupling.

Furthermore, we study the capabilities of the model in both \textit{interpolation} (in-domain), where it is trained across the full kinematic range, and \textit{extrapolation} (out-of-domain), where it is trained on a restricted region and must infer the behaviour towards the edges.

\begin{figure*}[!t]
  \centering
  \begin{subfigure}[b]{0.495\textwidth}
    \includegraphics[width=\linewidth]{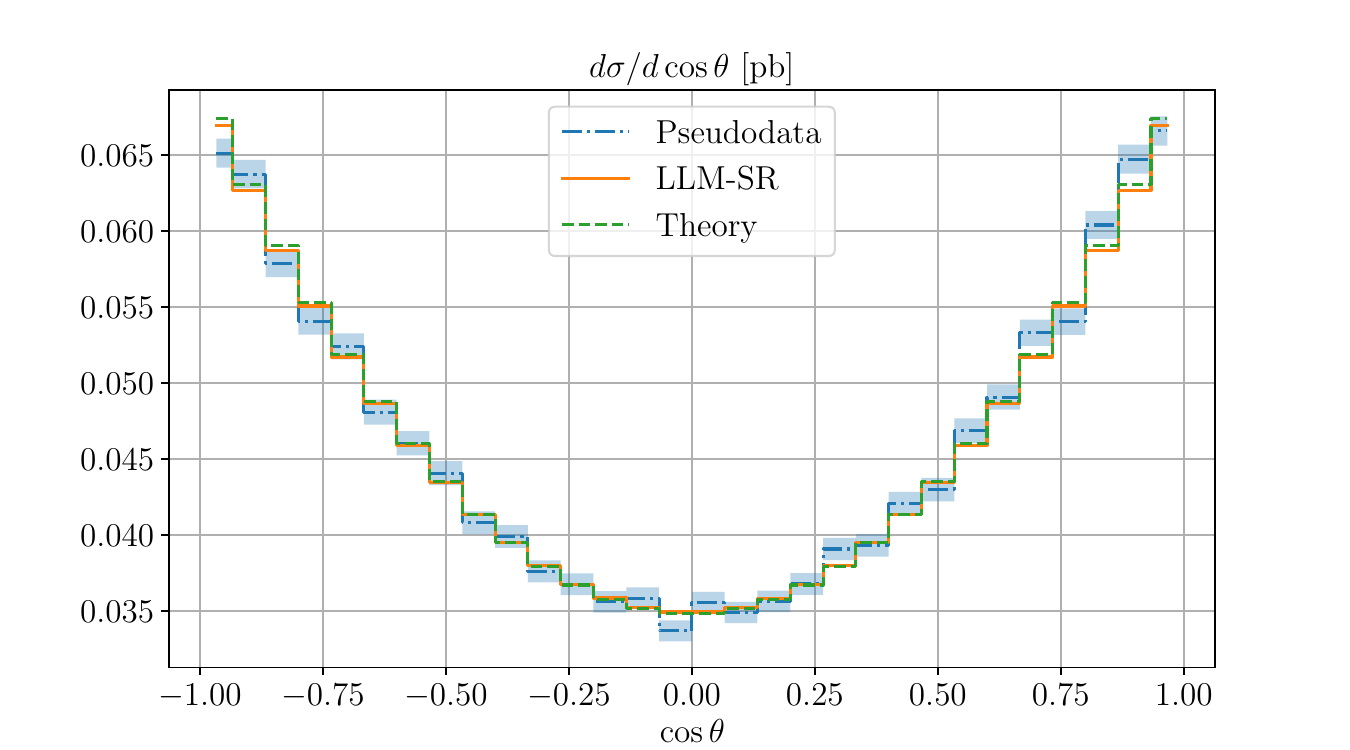}
  \end{subfigure}
  \begin{subfigure}[b]{0.495\textwidth}
    \includegraphics[width=\linewidth]{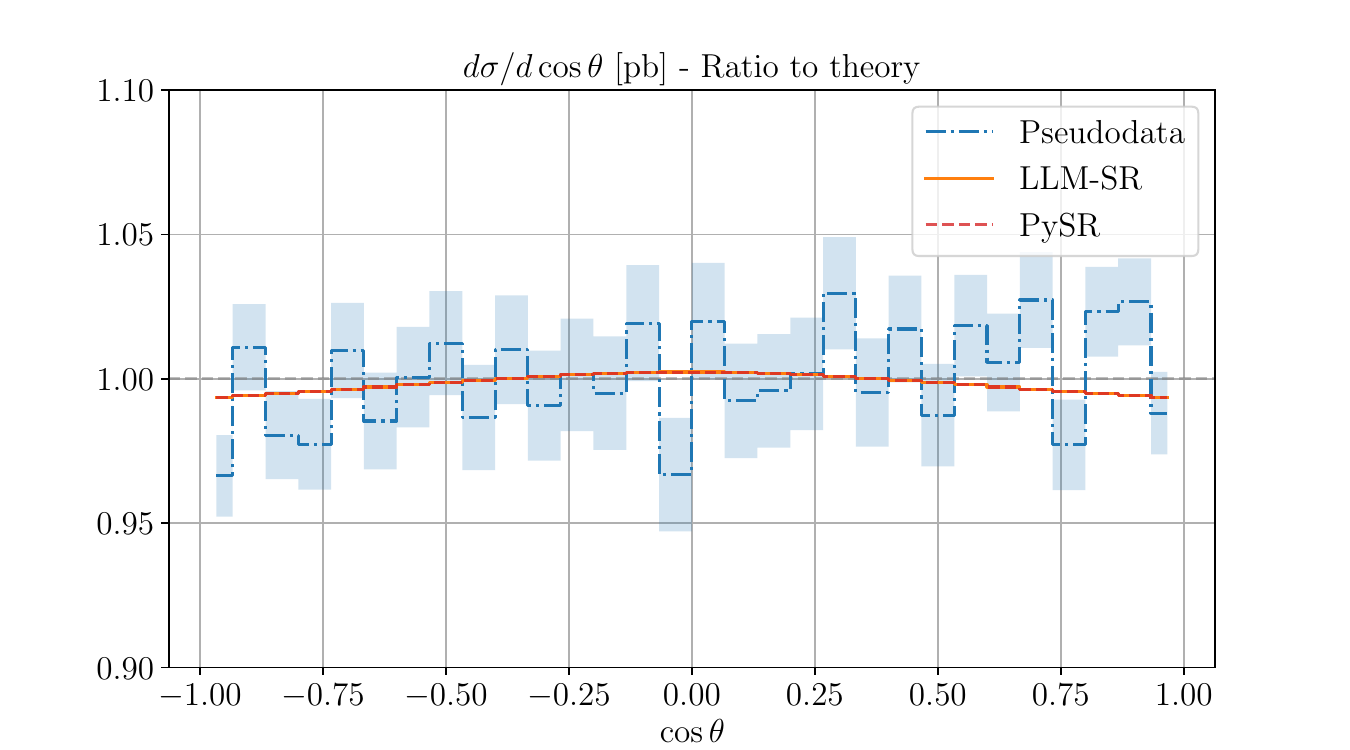}
  \end{subfigure}
  \caption{\justifying Lepton angular distribution. \textit{Left:} absolute differential cross section. \textit{Right:} ratio to the analytic formula.}
  \label{fig:lepton_interpolation}
\end{figure*}

Figure~\ref{fig:lepton_interpolation} shows the LLM-SR fit obtained using the full kinematic range. To assess interpolation capabilities, we train the regressor on a random $80\%$ subsample of the data and evaluate its performance across the domain. The LLM-SR prediction closely follows the analytic result, remaining within $\sim 1\%$ of the true distribution, even in bins where statistical fluctuations cause the pseudodata to deviate beyond their uncertainties. For comparison, the right panel also shows results from PySR~\cite{pysr}, a state-of-the-art, highly optimised SR algorithm based on multi-population evolutionary search over expression trees. 

Beyond fitting the distributions, LLM-SR also recovers the closed-form expression of Eq.~(\ref{eq:lepton}), derived from first-principles theory, as shown in Table~\ref{table:llmsr_vs_pysr_int}. As previously demonstrated in~\cite{Morales-Alvarado:2024jrk,Bendavid:2025urn}, PySR is likewise able to recover the correct expression. In fact, both methods identify the underlying equation with practically identical losses, defined as the weighted $\chi^2$ from squared residuals with weights $1/\sigma^2$ (where $\sigma$ denotes the bin uncertainty), normalised by their sum.

\begin{table}[ht]
\footnotesize
\setlength{\tabcolsep}{5pt}
\renewcommand{\arraystretch}{1.5}
\centering
\begin{tabular}{@{}c c c@{}}
\toprule
Method & Equation & Loss \\
\midrule
PySR & $f(\cos\theta) = 0.03492 + 0.03429 \cos^2\theta$ & $6.879 \cdot 10^{-7}$ \\
LLM-SR & $f(\cos\theta) = 0.03492 + 0.03428 \cos^2\theta$ & $6.881 \cdot 10^{-7}$ \\
\bottomrule
\end{tabular}
\caption{\justifying Comparison of PySR and LLM-SR equations in interpolation for the lepton angular distribution.}
\label{table:llmsr_vs_pysr_int}
\end{table}

\begin{figure}[!t]
  \centering
  \includegraphics[width=\linewidth]{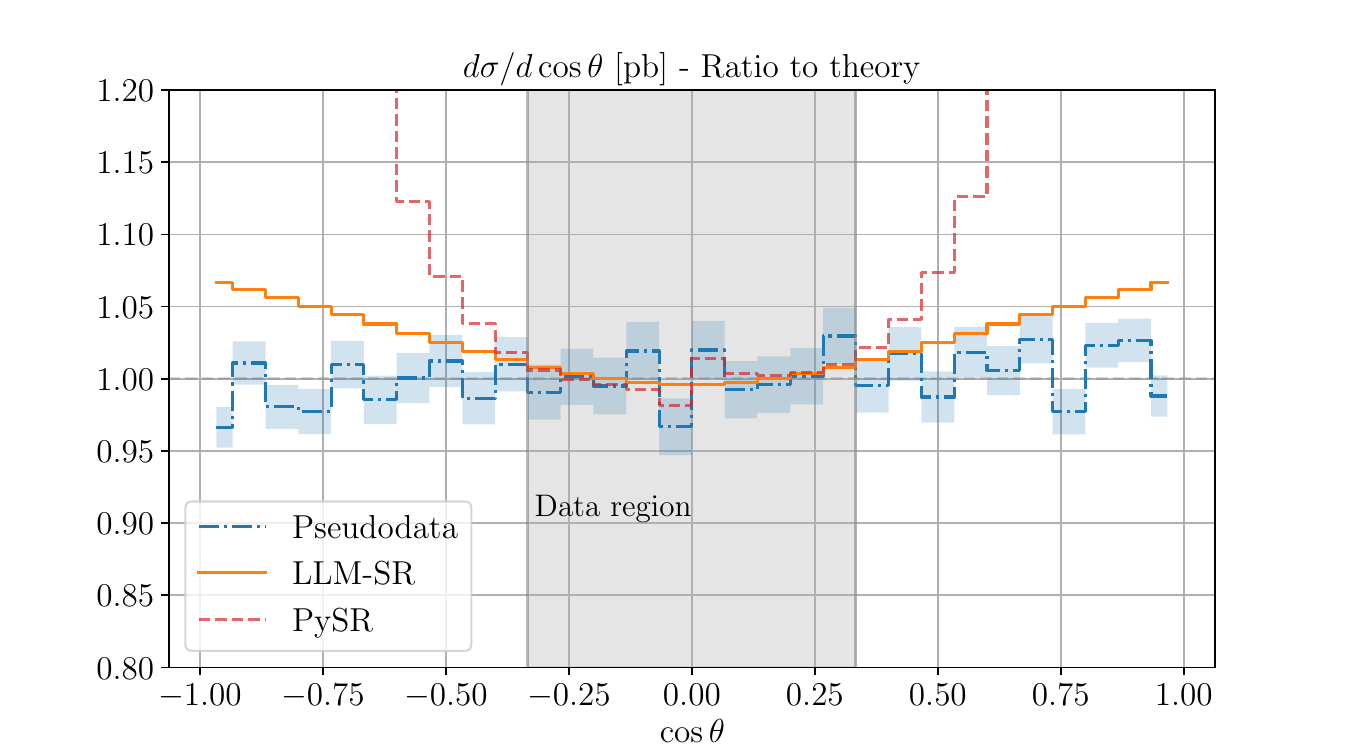}
  \caption{\justifying Ratio to the analytic formula, obtained when training only in the central region of $\cos\theta$ and extrapolating to the full domain.}
  \label{fig:lepton_extrapolation}
\end{figure}

We next investigate out-of-domain performance in extrapolation. In this case, the LLM-SR model is trained only on data from the central region around $\cos\theta \simeq 0$, and its predictions are evaluated across the full angular range. As shown in Figure~\ref{fig:lepton_extrapolation}, while both LLM-SR and PySR achieve comparable accuracy in the training region, their extrapolations differ markedly: LLM-SR remains closer to the true distribution, whereas PySR produces stronger deviations. This behaviour can be traced to the broader prior knowledge encoded in LLM-SR: the model is informed in the program that it is regressing an angular distribution (naturally, not being told about the exact functional form of the distribution), which guides it toward physically plausible extrapolations. By contrast, PySR, operating purely on the sample without external context, is more prone to overfitting within the training window, thus leading to poorer out-of-domain results.


Table~\ref{table:llmsr_vs_pysr_extra} reports the functional forms obtained by each method. LLM-SR successfully identifies the correct expression, while PySR converges to a more convoluted rational form that fits the in-domain training region but departs significantly from the underlying physics when extrapolated (this was confirmed in several independent runs). The losses, computed over the entire angular range, confirm the good out-of-domain generalisation of LLM-SR. Importantly, this behaviour illustrates how LLM-SR can leverage prior symbolic knowledge to recover physically meaningful equations even when trained on partial or noisy data. 

Of course, in this benchmark the true equation is known. However, even when the exact form is not available, the prior knowledge embedded in the LLM can guide the search towards compact, physically plausible expressions by drawing on related structures in its corpus.

\begin{table}[ht]
\footnotesize
\setlength{\tabcolsep}{5pt}
\renewcommand{\arraystretch}{1.5}
\centering
\begin{tabular}{@{}c c c@{}}
\toprule
Method & Equation & Loss \\
\midrule
PySR & $f(\cos\theta) = \dfrac{0.03399}{- \cos^2\theta + 0.97790 - \tfrac{0.00053}{\cos\theta}}$ & $2.417 \cdot 10^{-2}$ \\
LLM-SR & $f(\cos\theta) = 0.03469 + 0.03981 \cos^2\theta$ & $1.462 \cdot 10^{-6}$ \\
\bottomrule
\end{tabular}
\caption{\justifying Comparison of PySR and LLM-SR extrapolation models for the lepton angular distribution.}
\label{table:llmsr_vs_pysr_extra}
\end{table}

\section{Equation discovery}
\label{sec:discovery}

Having validated the ability of LLM-SR to reproduce known analytical laws from imperfect pseudodata, we now turn to the discovery of unknown closed-form expressions for the angular coefficients $A_i$ from a foundation model perspective. The angular coefficients play a crucial role in precision electroweak physics at the LHC~\cite{Mirkes:1992hu,Mirkes:1994dp,Mirkes:1994eb,Mirkes:1994nr}. In particular, they parametrise the full 5-dimensional differential cross section for for Drell–Yan vector boson production and decay as
{\small
\begin{equation}
\begin{aligned}
  \frac{d^5\sigma}{dp_T\, dy \, dm \, d\cos\theta \, d\phi}
   &= \frac{3}{16\pi}\,
      \frac{d^3\sigma^{U+L}}{dp_T\, dy\, dm} \\
   &\times \left[(1+\cos^2\theta) 
      + \sum_{i=0}^7 P_i(\theta,\phi)\,A_i \right],
\end{aligned}
\label{eq:full_coeff}
\end{equation}
}
with $p_T$ the transverse momentum of the $Z$ boson, $y$ the rapidity of the dilepton system, $m$ its invariant mass, $\theta$ and $\phi$ the polar and azimuthal angles of the lepton in the Collins–Soper frame~\cite{Collins:1977iv}, and the functions $P_i(\theta,\phi)$ are spherical harmonic polynomials. 

The $A_i$ coefficients, which encode the $p_T$, $y$, and $m$ dependence of the cross section, are not known in closed analytic form beyond specific asymptotic limits. They can instead be extracted by evaluating weighted averages over angular distributions obtained from simulations at fixed order~\cite{Bern:2011ie}. Compact, differentiable expressions for these quantities are therefore of substantial interest: they can accelerate phenomenological analyses, provide insight into scaling behaviours, e.g. the Lam–Tung relation~\cite{Lam:1978zr,Lam:1978pu,Lam:1980uc,Gehrmann-DeRidder:2015wbt,Gehrmann-DeRidder:2016jns,Gauld:2024glt,Piloneta:2024aac,Li:2024iyj,Li:2025fom,Arroyo-Castro:2025slx}, and allow for smooth extrapolation into kinematic regions where good fidelity simulation is not available. LLM-SR provides a novel avenue to obtain such parametrisations, with the added benefit of introducing domain knowledge if required.

In the following, we apply LLM-SR to obtain analytic approximations for selected angular coefficients, progressively increasing the dimensionality of the kinematic dependence. We begin with the 1D function $A_0(p_T)$, then extend to 2D distributions in $A_4(p_T, |y|)$, and finally consider the fully differential 3D case $A_4(p_T, |y|, m)$. 
The same LLM-SR methodology can in principle be applied to recover compact expressions for the full set of coefficients in any dimensionality. For a comprehensive collection of high-accuracy closed-form results obtained with the evolutionary methods of PySR, we refer the reader to Refs.~\cite{Morales-Alvarado:2024jrk,Bendavid:2025urn}.

\subsection{1D Angular Coefficients}

As a first step, we focus on $A_0(p_T)$, marginalising over rapidity and invariant mass. In this case we allow for 4 free parameters, leading the LLM-SR framework to identify a compact rational function:
\begin{equation}
\label{eq:a0_1d}
A_0(p_T) = 
\frac{\alpha_{2}\,p_T^{2} + \alpha_{3}\,p_T^{4}}
     {\alpha_{1} + \alpha_{4}\,p_T^{2} + p_T^{4}}\,,
\end{equation}
with numerical coefficients
$\alpha_{1}=3.88\times10^{1}$, 
$\alpha_{2}=-2.17\times10^{1}$, 
$\alpha_{3}=9.47\times10^{-1}$, 
$\alpha_{4}=2.83\times10^{3}$. 
The associated loss is $L = 1.892 \times 10^{-5}$. Note that, from now on, all dimensionful variables are understood to be normalised by their corresponding units so as to ensure the dimensional consistency of the expressions. The LLM-SR program used to find this expression is described in Program \ref{fig:program_1} in the Appendix, which shows the minimal LLM-SR setup used for this task. The docstring specifies the regression target and embeds domain knowledge (e.g. $A_0\!\to\!0$ as $p_T\!\to\!0$). The variable \texttt{MAX\_NPARAMS = 4} fixes the maximum number of free numerical parameters for the search. The routine \texttt{evaluate(\dots)} delegates loss computation to the user, permitting choices such as MSE or a weighted $\chi^2$. Finally, \texttt{equation\_v0(\dots)} provides a simple program prior: here a rational form $\propto p_T^2/(c+p_T^2)$ enforcing the known asymptotic limit as $p_T\!\to\!0$, which the LLM refines by proposing skeleton programs whose parameters are subsequently optimised.

\begin{figure}[!t]
  \centering
  \includegraphics[width=0.48\textwidth]{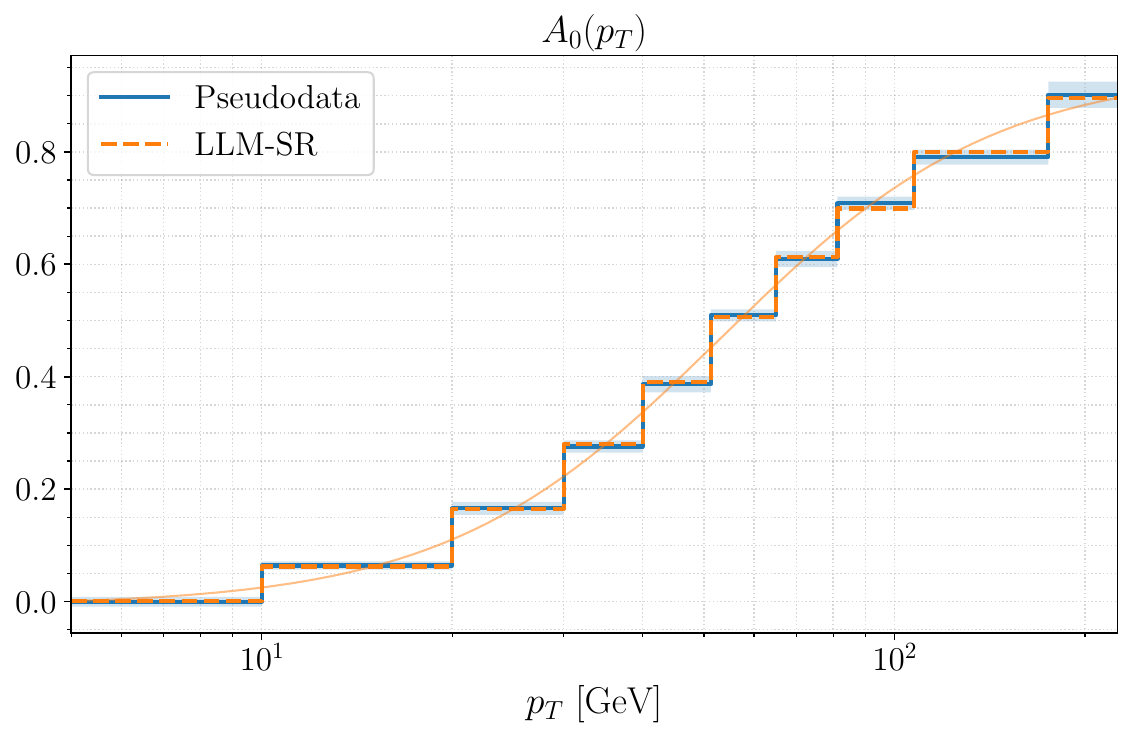}
  \caption{\justifying $A_0 (p_T)$ pseudodata from the simulation and LLM-SR result as per Eq. (\ref{eq:a0_1d}).}
  \label{fig:angular_1d}
\end{figure}

Figure~\ref{fig:angular_1d} shows the agreement between this expression and the pseudodata obtained from simulation.

The quality of the fit is comparable to the best results obtained with algorithms such as PySR~\cite{Bendavid:2025urn}. However, the LLM-SR solution presents an additional feature: it can be instructed to reproduce the correct collinear behaviour. In this case, the expression vanishes as $p_T \to 0$, consistent with the scaling derived from first-principles in the collinear limit. The reference PySR result approaches a small but finite value, an artefact of being trained on discrete bin centres that do not probe arbitrarily low transverse momenta. This built-in consistency with theoretical expectations makes the LLM-SR parametrisation more robust for extrapolation across the full kinematic range. In particular, it ensures reliable behaviour in regions where data are limited, while retaining compactness and interpretability.

\subsection{Methodological considerations}

\begin{figure}[!t]
  \centering
  \includegraphics[width=\linewidth]{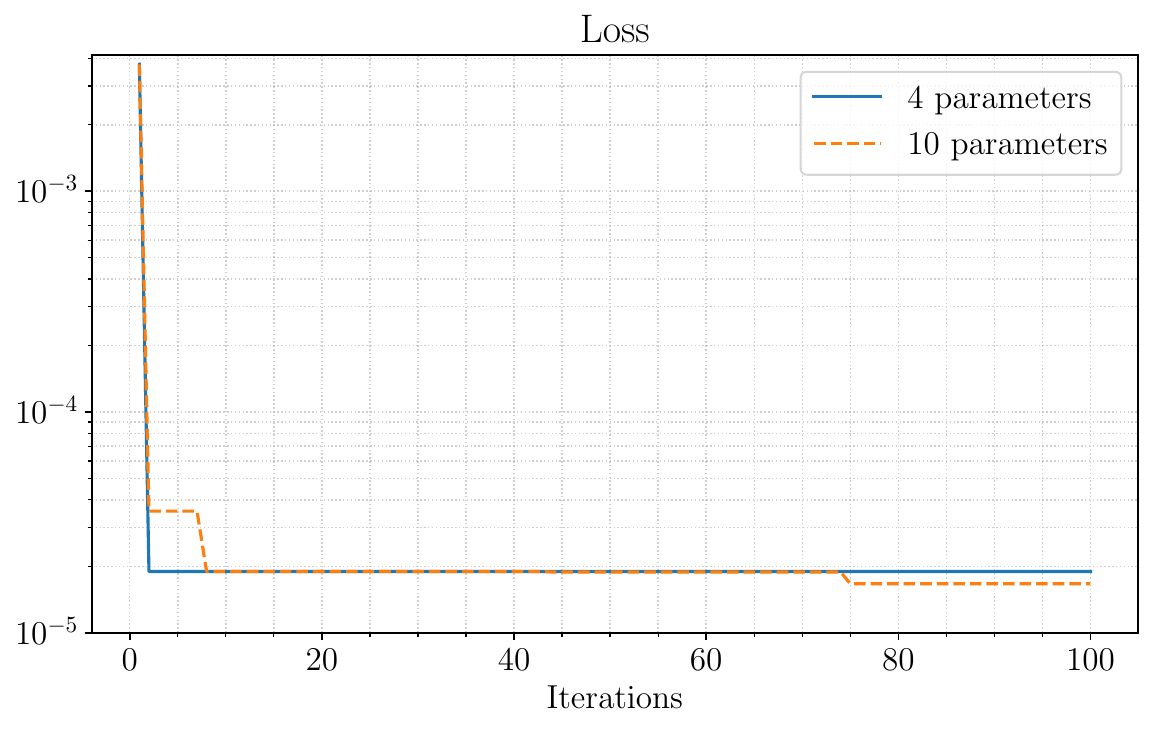}
  \caption{\justifying Loss trajectories for models with 4 and 10 free parameters.}
  \label{fig:loss_params}
\end{figure}

As discussed in the introduction, the behaviour of LLM-SR can be influenced by several factors. One such factor is the number of free parameters available to construct equations. Increasing the number of parameters naturally provides more flexibility to accommodate the dataset, often leading to smaller losses. Figure ~\ref{fig:loss_params} shows the loss trajectories for models with 4 and 10 free parameters, trained with the same program prior (the one of Program \ref{fig:program_1}). For most of the training iterations the performance is comparable, but once sufficient iterations have elapsed the model with more parameters achieves a slightly smaller loss. This behaviour reflects the trade-off between compactness and descriptive power: additional parameters can improve accuracy, but at the cost of reduced simplicity.

\begin{figure}[!t]
  \centering
  \includegraphics[width=\linewidth]{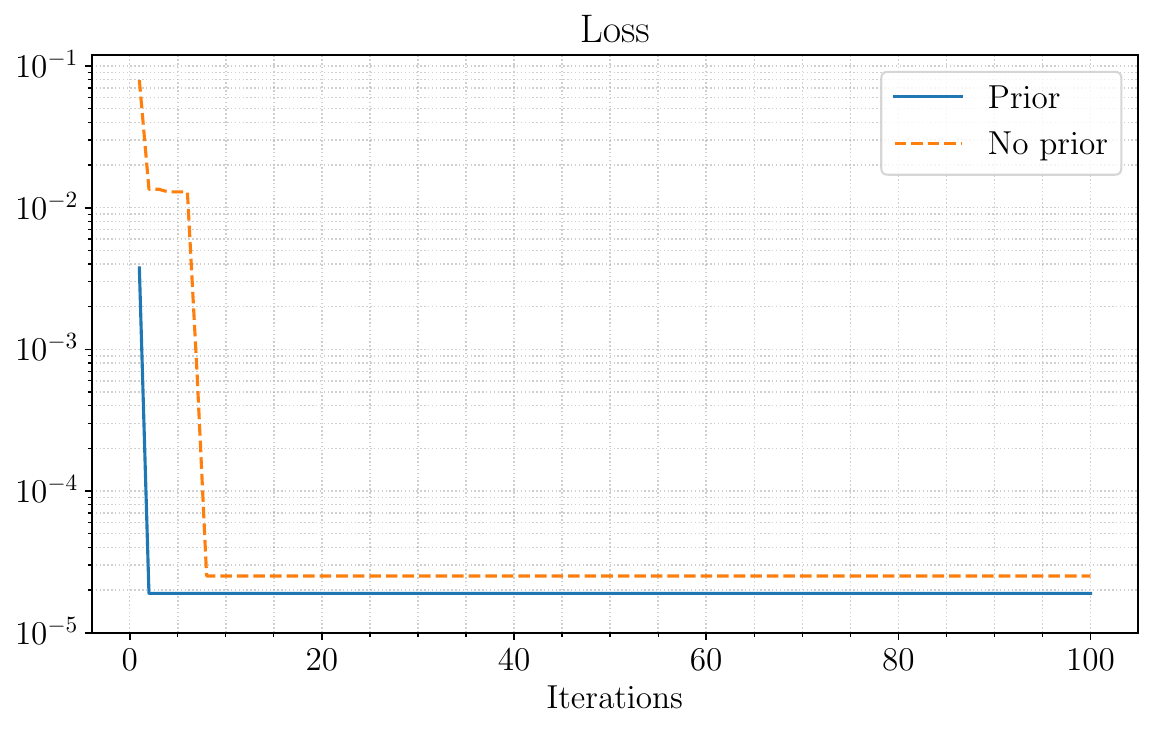}
  \caption{\justifying Loss trajectories for models with and without a program prior.}
  \label{fig:loss_prior}
\end{figure}

A second important consideration is the role of the program prior provided to the LLM at the start of training. Figure~\ref{fig:loss_prior} compares the loss evolution for models with the same number of parameters (4, in this case) but different priors. As expected, the model initialised with an informative program prior, \texttt{equation\_v0} in Program \ref{fig:program_1}, reaches better performance from the outset. Interestingly, however, the model without an informative prior, shown in Program \ref{fig:program_2} in the Appendix, continues to improve steadily during training, eventually approaching (but not surpassing) the performance of the model with informative prior. This indicates that priors, especially when they encode domain-specific knowledge, can accelerate convergence and yield consistently lower losses. These observations are consistent with the ablation study reported in Ref.~\cite{llmsr}, and highlight the dual role of priors in both guiding the search and regularising the solutions toward physically meaningful expressions.

\subsection{Multidimensional Angular Coefficients}

We now extend the analysis to 2D distributions, focusing on the coefficient $A_4(p_T, |y|)$. Compared to the 1D case, the inclusion of rapidity as a second variable provides additional structure that can reveal more distinctive and characteristic patterns in the regression.

\begin{figure*}[!t]
  \centering
  \includegraphics[width=0.8\textwidth]{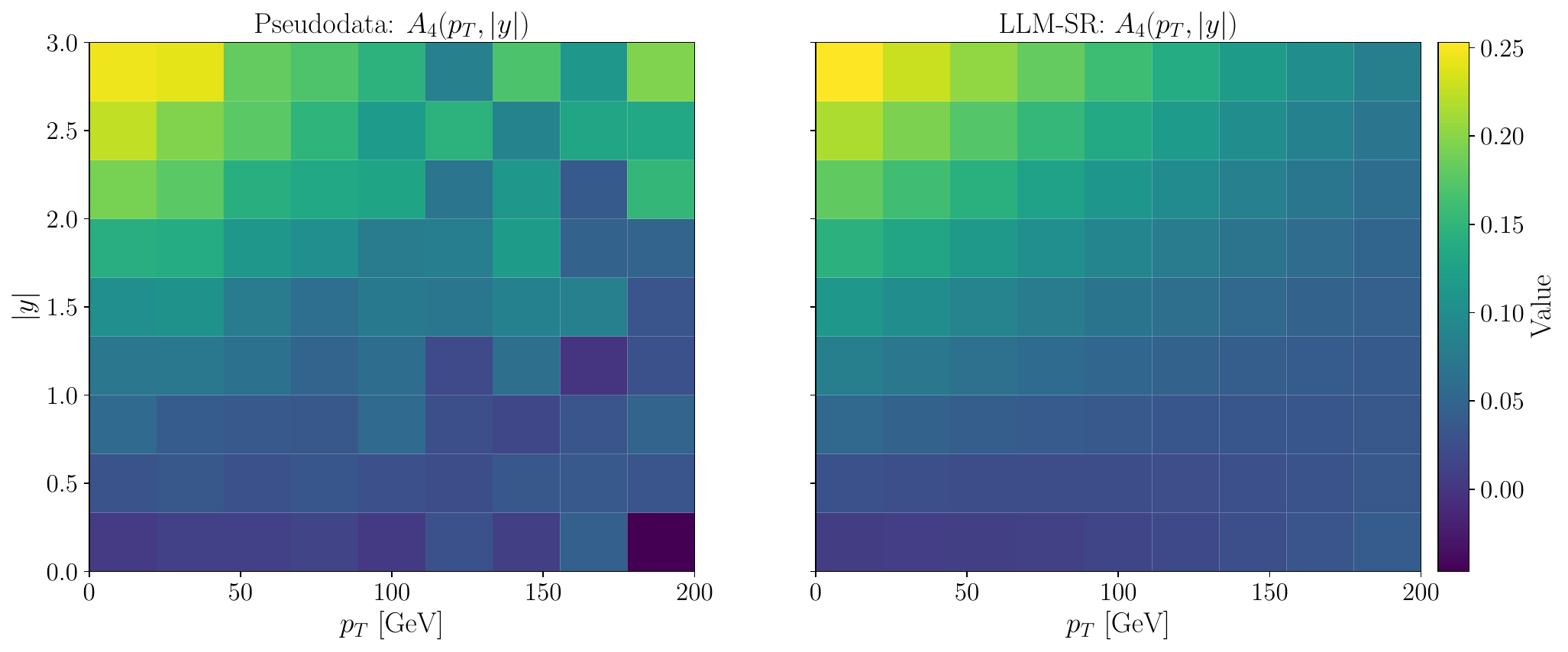}
  \caption{2D $A_4(p_T,|y|)$ in pseudodata from the simulation and LLM-SR.}
  \label{fig:angular_2d}
\end{figure*}

For this task, we constrain LLM-SR to 6 free numerical parameters, aiming a balance between compactness and descriptive power. The method yields smooth and differentiable expressions across the full kinematic domain, with the resulting functional form
\begin{equation}
\begin{aligned}
A_4(p_T,|y|) &= \beta_{1}\,p_T^{\beta_{2}} + \beta_{3}\,|y|^{\beta_{4}} \\
             &\quad + \beta_{5}\,p_T|y| + \beta_{6}\,(p_T^2+|y|^2)\,,
\end{aligned}
\label{eq:a4-ansatz}
\end{equation}
with parameters $\beta_{1}=2.11\times10^{-5}$, $\beta_{2}=1.19$, $\beta_{3}=7.13\times10^{-2}$, $\beta_{4}=1.27$, $\beta_{5}=-4.31\times10^{-4}$, $\beta_{6}=9.83\times10^{-7}$, and an associated loss $L = 1.316 \times 10^{-4}$. The performance of this expression compared to the pseudodata is shown in Fig.~\ref{fig:angular_2d}. The LLM-SR expression provides a more regular description at higher $p_T$, mitigating the distortions introduced by statistical fluctuations, which become more important in this region.

Interestingly, the achieved loss is comparable to the most accurate solutions in the “hall of fame” obtained with PySR in Ref.~\cite{Bendavid:2025urn}, which also finds similar functional forms with a comparable number of tunable parameters. This agreement is reassuring: despite the very different search strategies (LLM-guided symbolic priors versus genetic algorithms on expression trees), both methods converge to parametrisations of similar complexity and accuracy. 

\begin{figure*}[!t]
  \centering
  \includegraphics[width=0.8\textwidth]{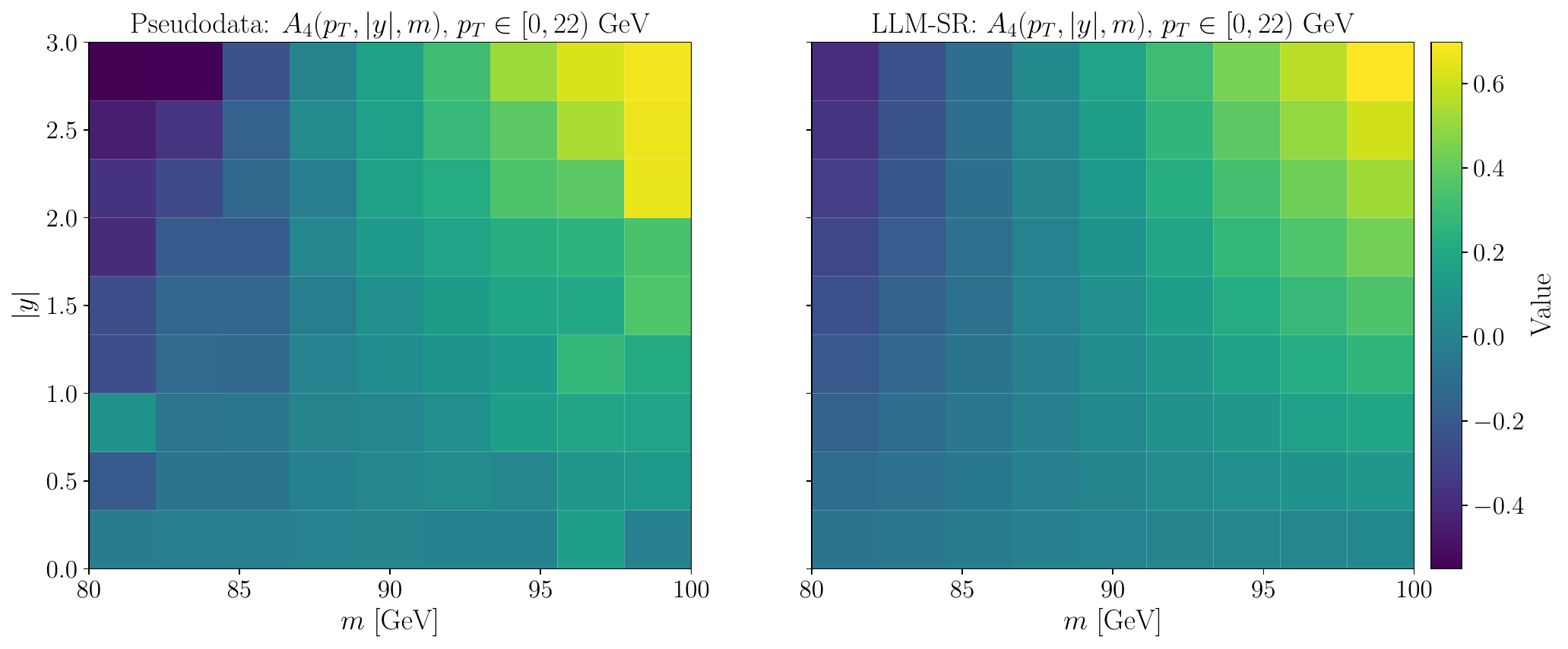}\\[1em]
  \includegraphics[width=0.8\textwidth]{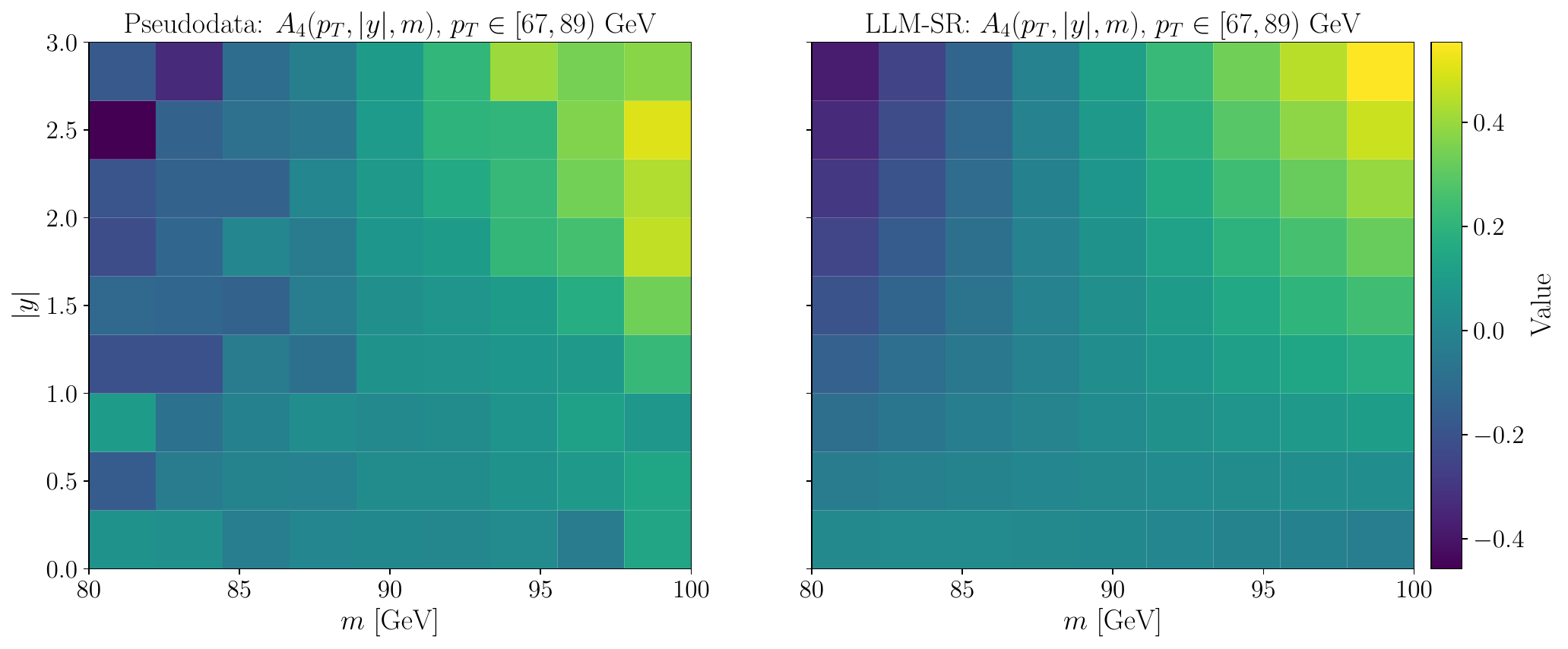}
  \caption{\justifying 3D $A_4(p_T,|y|,m)$. 
  Top panel: comparison between pseudodata and the LLM-SR prediction in the bin $p_T \in [0, 22)\,\text{GeV}$. 
  Bottom panel: same comparison in the bin $p_T \in [67, 89)\,\text{GeV}$.}
  \label{fig:angular_3d}
\end{figure*}

Finally, we turn to the fully differential case and study the 3D dependence of the angular coefficient $A_4(p_T,|y|,m)$. For this task, we allow LLM-SR to use 10 free numerical parameters. The method identifies a simple quadratic polynomial structure supplemented by mixed interference terms:
\begin{equation}
\begin{aligned}
A_4(p_T,|y|,m) &= \gamma_{1}
+ \gamma_{2}\,p_T
+ \gamma_{3}\,|y|
+ \gamma_{4}\,m  \\
&\quad + \gamma_{5}\,p_T^{2}
+ \gamma_{6}\,|y|^{2}
+ \gamma_{7}\,m^{2}  \\
&\quad + \gamma_{8}\,p_T|y|
+ \gamma_{9}\,p_T m
+ \gamma_{10}\,|y|m\,,
\end{aligned}
\label{eq:a4-3d-ansatz}
\end{equation}
with parameters
$\gamma_{1}=-2.47$, 
$\gamma_{2}=1.19\times10^{-2}$, 
$\gamma_{3}=-1.83$, 
$\gamma_{4}=5.10\times10^{-2}$, 
$\gamma_{5}=1.46\times10^{-6}$, 
$\gamma_{6}=9.14\times10^{-3}$, 
$\gamma_{7}=-2.62\times10^{-4}$, 
$\gamma_{8}=-4.39\times10^{-4}$, 
$\gamma_{9}=-1.31\times10^{-4}$, 
$\gamma_{10}=2.08\times10^{-2}$, 
and an associated loss $L = 1.383 \times 10^{-3}$. 
The comparison with pseudodata is shown in Fig.~\ref{fig:angular_3d} for two representative transverse-momentum slices: 
$p_T \in [0, 22)\,\text{GeV}$ and $p_T \in [67, 89)\,\text{GeV}$.

As in the 1D and 2D cases, the LLM-SR ansatz yields a compact, differentiable representation that captures the global trends of the simulation while mitigating statistical noise. In particular, it provides a more regular and stable description in regions where the pseudodata are most affected by fluctuations, such as at invariant masses far away from the $Z$ pole. This illustrates how LLM-guided SR can generalise to higher-dimensional settings, retaining interpretability while achieving accuracy comparable to that of other approaches~\cite{Bendavid:2025urn}.

\section{Conclusions}
\label{sec:conclusions}

We have investigated the use of foundation models, exemplified by LLM-SR,, for equation discovery in HEP. In a controlled validation with lepton angular distributions, LLM-SR successfully recovered the known analytic law. Its in-domain accuracy matched that of state-of-the-art baselines, providing a valuable cross-check of methodological consistency. In applications to angular coefficients in Drell–Yan production, LLM-SR produced compact parametrisations with accuracy comparable to existing methods. In addition, the approach exhibits appealing features, such as more stable behaviour when extrapolating beyond the training domain and the ability to explore a broader range of functional forms without the need to predefine an operator basis. These advantages are linked to the incorporation of domain knowledge and scientific priors into the program generation process, which helps guide the search towards physically meaningful structures.

Although domain knowledge does not yield dramatic improvements in quantitative performance, informative prompts and program priors can steer the search towards expressions with more natural physical properties (e.g. correct asymptotics), thereby enhancing interpretability and robustness, particularly in out-of-domain regions. LLMs thus provide an efficient mechanism to navigate the combinatorial space of candidate equations and propose viable structures guided by embedded scientific priors.

There remain aspects that would benefit from further development. For the problems considered, LLM-SR typically requires somewhat longer runtimes than evolutionary SR methods (e.g. PySR), and its performance is influenced by the choice of hyperparameters (e.g. number of free parameters) together with prompt and program design. Looking ahead, scaling to additional observables could benefit from more automated hyperparameter selection, systematic program prior construction, and improved use of experience buffers. From a phenomenological perspective, future studies should also incorporate additional sources of systematic uncertainty and correlations.

As exemplified by LLMs, foundation models offer a promising avenue for equation discovery in HEP. Their ability to combine data-driven modelling with embedded scientific priors makes them a valuable complement to existing approaches and a practical bridge towards compact, interpretable, and physics-aware analytical forms that can aid discovery in fundamental physics.

\newpage
\section*{Appendix}
\label{app:extra}

This appendix provides supplementary material to illustrate the practical implementation of the LLM-SR methodology. As an example, Program ~\ref{fig:program_1} shows a minimal Python program used to obtain $A_0(p_T)$. The docstring encodes the task description and physical constraints, while the functions specify the evaluation of the loss and a simple program prior for the regression. The non-informative prior is given in Program ~\ref{fig:program_2}.

\begin{figure*}[!ht]
\captionsetup{skip=0pt} 
\centering
\begin{lstlisting}[language=Python, caption={LLM-SR Python program to obtain $A_0(p_T)$.}, label={fig:program_1}]
"""
- Find the mathematical equation that describes this angular coefficient in high energy particle collisions.
- The specific angular coefficient is A0, and you have to express it as A0(pt_centres).
- You are given the value of bin centres in transverse momentum pT, the value of the angular coefficient in that bin, and its corresponding uncertainty.
- In the limit of pt_centres going to zero, the value has to vanish exactly. Respect this strictly.
"""

# Number of parameters
MAX_NPARAMS = 4

def evaluate(data, equation):
    """
    Evaluate the equation on data observations.
    """
    ...

def equation_v0(pt_centres, params):
    """ Mathematical function for the angular coefficient.

    Args:
        pt_centres: A numpy array representing a set of pt centres in a distribution in particle physics.
        params: Array of numeric constants or parameters to be optimized

    Return:
        A numpy array representing the value of the angular coefficient as a function of pT in each bin.
    """
    return pt_centres**2 / (params[0] + pt_centres**2)
    
\end{lstlisting}
\end{figure*}

\begin{figure*}[!ht]
\captionsetup{skip=0pt} 
\centering
\begin{lstlisting}[language=Python, caption={Same as Program \ref{fig:program_1} but with a flat, non-informative \texttt{equation\_v0} prior.}, label={fig:program_2}]
def equation_v0(pt_centres, params):
    """ Mathematical function for the angular coefficient.

    Args:
        pt_centres: A numpy array representing a set of pt centres in a distribution in particle physics.
        params: Array of numeric constants or parameters to be optimized

    Return:
        A numpy array representing the value of the angular coefficient as a function of pT in each bin.
    """
    return params[0]    
\end{lstlisting}
\end{figure*}

\bibliography{references}

\begin{thebibliography}{56}%
\makeatletter
\providecommand \@ifxundefined [1]{%
 \@ifx{#1\undefined}
}%
\providecommand \@ifnum [1]{%
 \ifnum #1\expandafter \@firstoftwo
 \else \expandafter \@secondoftwo
 \fi
}%
\providecommand \@ifx [1]{%
 \ifx #1\expandafter \@firstoftwo
 \else \expandafter \@secondoftwo
 \fi
}%
\providecommand \natexlab [1]{#1}%
\providecommand \enquote  [1]{``#1''}%
\providecommand \bibnamefont  [1]{#1}%
\providecommand \bibfnamefont [1]{#1}%
\providecommand \citenamefont [1]{#1}%
\providecommand \href@noop [0]{\@secondoftwo}%
\providecommand \href [0]{\begingroup \@sanitize@url \@href}%
\providecommand \@href[1]{\@@startlink{#1}\@@href}%
\providecommand \@@href[1]{\endgroup#1\@@endlink}%
\providecommand \@sanitize@url [0]{\catcode `\\12\catcode `\$12\catcode
  `\&12\catcode `\#12\catcode `\^12\catcode `\_12\catcode `\%12\relax}%
\providecommand \@@startlink[1]{}%
\providecommand \@@endlink[0]{}%
\providecommand \url  [0]{\begingroup\@sanitize@url \@url }%
\providecommand \@url [1]{\endgroup\@href {#1}{\urlprefix }}%
\providecommand \urlprefix  [0]{URL }%
\providecommand \Eprint [0]{\href }%
\providecommand \doibase [0]{http://dx.doi.org/}%
\providecommand \selectlanguage [0]{\@gobble}%
\providecommand \bibinfo  [0]{\@secondoftwo}%
\providecommand \bibfield  [0]{\@secondoftwo}%
\providecommand \translation [1]{[#1]}%
\providecommand \BibitemOpen [0]{}%
\providecommand \bibitemStop [0]{}%
\providecommand \bibitemNoStop [0]{.\EOS\space}%
\providecommand \EOS [0]{\spacefactor3000\relax}%
\providecommand \BibitemShut  [1]{\csname bibitem#1\endcsname}%
\let\auto@bib@innerbib\@empty
\bibitem [{\citenamefont {Bommasani}(2021)}]{bommasani2021opportunities}%
  \BibitemOpen
  \bibfield  {author} {\bibinfo {author} {\bibfnamefont {R.}~\bibnamefont
  {Bommasani}},\ }\href@noop {} {\bibfield  {journal} {\bibinfo  {journal}
  {arXiv preprint arXiv:2108.07258}\ } (\bibinfo {year} {2021})}\BibitemShut
  {NoStop}%
\bibitem [{\citenamefont {Wei}\ \emph {et~al.}(2022)\citenamefont {Wei},
  \citenamefont {Tay}, \citenamefont {Bommasani}, \citenamefont {Raffel},
  \citenamefont {Zoph}, \citenamefont {Borgeaud}, \citenamefont {Yogatama},
  \citenamefont {Bosma}, \citenamefont {Zhou}, \citenamefont {Metzler} \emph
  {et~al.}}]{wei2022emergent}%
  \BibitemOpen
  \bibfield  {author} {\bibinfo {author} {\bibfnamefont {J.}~\bibnamefont
  {Wei}}, \bibinfo {author} {\bibfnamefont {Y.}~\bibnamefont {Tay}}, \bibinfo
  {author} {\bibfnamefont {R.}~\bibnamefont {Bommasani}}, \bibinfo {author}
  {\bibfnamefont {C.}~\bibnamefont {Raffel}}, \bibinfo {author} {\bibfnamefont
  {B.}~\bibnamefont {Zoph}}, \bibinfo {author} {\bibfnamefont {S.}~\bibnamefont
  {Borgeaud}}, \bibinfo {author} {\bibfnamefont {D.}~\bibnamefont {Yogatama}},
  \bibinfo {author} {\bibfnamefont {M.}~\bibnamefont {Bosma}}, \bibinfo
  {author} {\bibfnamefont {D.}~\bibnamefont {Zhou}}, \bibinfo {author}
  {\bibfnamefont {D.}~\bibnamefont {Metzler}},  \emph {et~al.},\ }\href@noop {}
  {\bibfield  {journal} {\bibinfo  {journal} {arXiv preprint arXiv:2206.07682}\
  } (\bibinfo {year} {2022})}\BibitemShut {NoStop}%
\bibitem [{\citenamefont {Yamada}\ \emph {et~al.}(2025)\citenamefont {Yamada},
  \citenamefont {Lange}, \citenamefont {Lu}, \citenamefont {Hu}, \citenamefont
  {Lu}, \citenamefont {Foerster}, \citenamefont {Clune},\ and\ \citenamefont
  {Ha}}]{yamada2025aiscientistv2workshoplevelautomated}%
  \BibitemOpen
  \bibfield  {author} {\bibinfo {author} {\bibfnamefont {Y.}~\bibnamefont
  {Yamada}}, \bibinfo {author} {\bibfnamefont {R.~T.}\ \bibnamefont {Lange}},
  \bibinfo {author} {\bibfnamefont {C.}~\bibnamefont {Lu}}, \bibinfo {author}
  {\bibfnamefont {S.}~\bibnamefont {Hu}}, \bibinfo {author} {\bibfnamefont
  {C.}~\bibnamefont {Lu}}, \bibinfo {author} {\bibfnamefont {J.}~\bibnamefont
  {Foerster}}, \bibinfo {author} {\bibfnamefont {J.}~\bibnamefont {Clune}}, \
  and\ \bibinfo {author} {\bibfnamefont {D.}~\bibnamefont {Ha}},\ }\href
  {https://arxiv.org/abs/2504.08066} {\enquote {\bibinfo {title} {The ai
  scientist-v2: Workshop-level automated scientific discovery via agentic tree
  search},}\ } (\bibinfo {year} {2025}),\ \Eprint
  {http://arxiv.org/abs/2504.08066} {arXiv:2504.08066 [cs.AI]} \BibitemShut
  {NoStop}%
\bibitem [{\citenamefont {Heneka}\ \emph {et~al.}(2025)\citenamefont {Heneka},
  \citenamefont {Nieser}, \citenamefont {Ore}, \citenamefont {Plehn},\ and\
  \citenamefont {Schiller}}]{Heneka:2025fpe}%
  \BibitemOpen
  \bibfield  {author} {\bibinfo {author} {\bibfnamefont {C.}~\bibnamefont
  {Heneka}}, \bibinfo {author} {\bibfnamefont {F.}~\bibnamefont {Nieser}},
  \bibinfo {author} {\bibfnamefont {A.}~\bibnamefont {Ore}}, \bibinfo {author}
  {\bibfnamefont {T.}~\bibnamefont {Plehn}}, \ and\ \bibinfo {author}
  {\bibfnamefont {D.}~\bibnamefont {Schiller}},\ }\href@noop {} {\  (\bibinfo
  {year} {2025})},\ \Eprint {http://arxiv.org/abs/2506.14757} {arXiv:2506.14757
  [astro-ph.CO]} \BibitemShut {NoStop}%
\bibitem [{\citenamefont {Kaiser}\ \emph {et~al.}(2024)\citenamefont {Kaiser},
  \citenamefont {Eichler},\ and\ \citenamefont {Lauscher}}]{Kaiser:2024lkg}%
  \BibitemOpen
  \bibfield  {author} {\bibinfo {author} {\bibfnamefont {J.}~\bibnamefont
  {Kaiser}}, \bibinfo {author} {\bibfnamefont {A.}~\bibnamefont {Eichler}}, \
  and\ \bibinfo {author} {\bibfnamefont {A.}~\bibnamefont {Lauscher}},\
  }\href@noop {} {\  (\bibinfo {year} {2024})},\ \Eprint
  {http://arxiv.org/abs/2405.08888} {arXiv:2405.08888 [cs.CL]} \BibitemShut
  {NoStop}%
\bibitem [{\citenamefont {Birk}\ \emph {et~al.}(2024)\citenamefont {Birk},
  \citenamefont {Hallin},\ and\ \citenamefont {Kasieczka}}]{Birk:2024knn}%
  \BibitemOpen
  \bibfield  {author} {\bibinfo {author} {\bibfnamefont {J.}~\bibnamefont
  {Birk}}, \bibinfo {author} {\bibfnamefont {A.}~\bibnamefont {Hallin}}, \ and\
  \bibinfo {author} {\bibfnamefont {G.}~\bibnamefont {Kasieczka}},\ }\href
  {\doibase 10.1088/2632-2153/ad66ad} {\bibfield  {journal} {\bibinfo
  {journal} {Mach. Learn. Sci. Tech.}\ }\textbf {\bibinfo {volume} {5}},\
  \bibinfo {pages} {035031} (\bibinfo {year} {2024})},\ \Eprint
  {http://arxiv.org/abs/2403.05618} {arXiv:2403.05618 [hep-ph]} \BibitemShut
  {NoStop}%
\bibitem [{\citenamefont {Qu}\ \emph {et~al.}(2022)\citenamefont {Qu},
  \citenamefont {Li},\ and\ \citenamefont {Qian}}]{Qu:2022mxj}%
  \BibitemOpen
  \bibfield  {author} {\bibinfo {author} {\bibfnamefont {H.}~\bibnamefont
  {Qu}}, \bibinfo {author} {\bibfnamefont {C.}~\bibnamefont {Li}}, \ and\
  \bibinfo {author} {\bibfnamefont {S.}~\bibnamefont {Qian}},\ }\href@noop {}
  {\  (\bibinfo {year} {2022})},\ \Eprint {http://arxiv.org/abs/2202.03772}
  {arXiv:2202.03772 [hep-ph]} \BibitemShut {NoStop}%
\bibitem [{\citenamefont {Golling}\ \emph {et~al.}(2024)\citenamefont
  {Golling}, \citenamefont {Heinrich}, \citenamefont {Kagan}, \citenamefont
  {Klein}, \citenamefont {Leigh}, \citenamefont {Osadchy},\ and\ \citenamefont
  {Raine}}]{Golling:2024abg}%
  \BibitemOpen
  \bibfield  {author} {\bibinfo {author} {\bibfnamefont {T.}~\bibnamefont
  {Golling}}, \bibinfo {author} {\bibfnamefont {L.}~\bibnamefont {Heinrich}},
  \bibinfo {author} {\bibfnamefont {M.}~\bibnamefont {Kagan}}, \bibinfo
  {author} {\bibfnamefont {S.}~\bibnamefont {Klein}}, \bibinfo {author}
  {\bibfnamefont {M.}~\bibnamefont {Leigh}}, \bibinfo {author} {\bibfnamefont
  {M.}~\bibnamefont {Osadchy}}, \ and\ \bibinfo {author} {\bibfnamefont
  {J.~A.}\ \bibnamefont {Raine}},\ }\href {\doibase 10.1088/2632-2153/ad64a8}
  {\bibfield  {journal} {\bibinfo  {journal} {Mach. Learn. Sci. Tech.}\
  }\textbf {\bibinfo {volume} {5}},\ \bibinfo {pages} {035074} (\bibinfo {year}
  {2024})},\ \Eprint {http://arxiv.org/abs/2401.13537} {arXiv:2401.13537
  [hep-ph]} \BibitemShut {NoStop}%
\bibitem [{\citenamefont {Harris}\ \emph {et~al.}(2025)\citenamefont {Harris},
  \citenamefont {Krupa}, \citenamefont {Kagan}, \citenamefont {Maier},\ and\
  \citenamefont {Woodward}}]{Harris:2024sra}%
  \BibitemOpen
  \bibfield  {author} {\bibinfo {author} {\bibfnamefont {P.}~\bibnamefont
  {Harris}}, \bibinfo {author} {\bibfnamefont {J.}~\bibnamefont {Krupa}},
  \bibinfo {author} {\bibfnamefont {M.}~\bibnamefont {Kagan}}, \bibinfo
  {author} {\bibfnamefont {B.}~\bibnamefont {Maier}}, \ and\ \bibinfo {author}
  {\bibfnamefont {N.}~\bibnamefont {Woodward}},\ }\href {\doibase
  10.1103/PhysRevD.111.032010} {\bibfield  {journal} {\bibinfo  {journal}
  {Phys. Rev. D}\ }\textbf {\bibinfo {volume} {111}},\ \bibinfo {pages}
  {032010} (\bibinfo {year} {2025})},\ \Eprint
  {http://arxiv.org/abs/2403.07066} {arXiv:2403.07066 [hep-ph]} \BibitemShut
  {NoStop}%
\bibitem [{\citenamefont {Mikuni}\ and\ \citenamefont
  {Nachman}(2025)}]{Mikuni:2024qsr}%
  \BibitemOpen
  \bibfield  {author} {\bibinfo {author} {\bibfnamefont {V.}~\bibnamefont
  {Mikuni}}\ and\ \bibinfo {author} {\bibfnamefont {B.}~\bibnamefont
  {Nachman}},\ }\href {\doibase 10.1103/PhysRevD.111.L051504} {\bibfield
  {journal} {\bibinfo  {journal} {Phys. Rev. D}\ }\textbf {\bibinfo {volume}
  {111}},\ \bibinfo {pages} {L051504} (\bibinfo {year} {2025})},\ \Eprint
  {http://arxiv.org/abs/2404.16091} {arXiv:2404.16091 [hep-ph]} \BibitemShut
  {NoStop}%
\bibitem [{\citenamefont {Ho}\ \emph {et~al.}(2024)\citenamefont {Ho},
  \citenamefont {Roberts}, \citenamefont {Han},\ and\ \citenamefont
  {Wang}}]{Ho:2024qyf}%
  \BibitemOpen
  \bibfield  {author} {\bibinfo {author} {\bibfnamefont {J.}~\bibnamefont
  {Ho}}, \bibinfo {author} {\bibfnamefont {B.~R.}\ \bibnamefont {Roberts}},
  \bibinfo {author} {\bibfnamefont {S.}~\bibnamefont {Han}}, \ and\ \bibinfo
  {author} {\bibfnamefont {H.}~\bibnamefont {Wang}},\ }\href@noop {} {\
  (\bibinfo {year} {2024})},\ \Eprint {http://arxiv.org/abs/2412.10665}
  {arXiv:2412.10665 [hep-ph]} \BibitemShut {NoStop}%
\bibitem [{\citenamefont {Bardhan}\ \emph {et~al.}(2025)\citenamefont
  {Bardhan}, \citenamefont {Agrawal}, \citenamefont {Tilak}, \citenamefont
  {Neeraj},\ and\ \citenamefont {Mitra}}]{Bardhan:2025icr}%
  \BibitemOpen
  \bibfield  {author} {\bibinfo {author} {\bibfnamefont {J.}~\bibnamefont
  {Bardhan}}, \bibinfo {author} {\bibfnamefont {R.}~\bibnamefont {Agrawal}},
  \bibinfo {author} {\bibfnamefont {A.}~\bibnamefont {Tilak}}, \bibinfo
  {author} {\bibfnamefont {C.}~\bibnamefont {Neeraj}}, \ and\ \bibinfo {author}
  {\bibfnamefont {S.}~\bibnamefont {Mitra}},\ }\href@noop {} {\  (\bibinfo
  {year} {2025})},\ \Eprint {http://arxiv.org/abs/2502.03933} {arXiv:2502.03933
  [cs.LG]} \BibitemShut {NoStop}%
\bibitem [{\citenamefont {Hallin}(2025)}]{Hallin:2025ywf}%
  \BibitemOpen
  \bibfield  {author} {\bibinfo {author} {\bibfnamefont {A.}~\bibnamefont
  {Hallin}}\ }(\bibinfo {year} {2025})\ \Eprint
  {http://arxiv.org/abs/2509.21434} {arXiv:2509.21434 [hep-ph]} \BibitemShut
  {NoStop}%
\bibitem [{\citenamefont {Petersen}\ \emph {et~al.}(2019)\citenamefont
  {Petersen}, \citenamefont {Landajuela}, \citenamefont {Mundhenk},
  \citenamefont {Santiago}, \citenamefont {Kim},\ and\ \citenamefont
  {Kim}}]{petersen2019deep}%
  \BibitemOpen
  \bibfield  {author} {\bibinfo {author} {\bibfnamefont {B.~K.}\ \bibnamefont
  {Petersen}}, \bibinfo {author} {\bibfnamefont {M.}~\bibnamefont
  {Landajuela}}, \bibinfo {author} {\bibfnamefont {T.~N.}\ \bibnamefont
  {Mundhenk}}, \bibinfo {author} {\bibfnamefont {C.~P.}\ \bibnamefont
  {Santiago}}, \bibinfo {author} {\bibfnamefont {S.~K.}\ \bibnamefont {Kim}}, \
  and\ \bibinfo {author} {\bibfnamefont {J.~T.}\ \bibnamefont {Kim}},\
  }\href@noop {} {\bibfield  {journal} {\bibinfo  {journal} {arXiv preprint
  arXiv:1912.04871}\ } (\bibinfo {year} {2019})}\BibitemShut {NoStop}%
\bibitem [{\citenamefont {Mundhenk}\ \emph {et~al.}(2021)\citenamefont
  {Mundhenk}, \citenamefont {Landajuela}, \citenamefont {Glatt}, \citenamefont
  {Santiago}, \citenamefont {Petersen} \emph {et~al.}}]{mundhenk2021symbolic}%
  \BibitemOpen
  \bibfield  {author} {\bibinfo {author} {\bibfnamefont {T.}~\bibnamefont
  {Mundhenk}}, \bibinfo {author} {\bibfnamefont {M.}~\bibnamefont
  {Landajuela}}, \bibinfo {author} {\bibfnamefont {R.}~\bibnamefont {Glatt}},
  \bibinfo {author} {\bibfnamefont {C.~P.}\ \bibnamefont {Santiago}}, \bibinfo
  {author} {\bibfnamefont {B.~K.}\ \bibnamefont {Petersen}},  \emph {et~al.},\
  }\href@noop {} {\bibfield  {journal} {\bibinfo  {journal} {Advances in Neural
  Information Processing Systems}\ }\textbf {\bibinfo {volume} {34}},\ \bibinfo
  {pages} {24912} (\bibinfo {year} {2021})}\BibitemShut {NoStop}%
\bibitem [{\citenamefont {Udrescu}\ and\ \citenamefont
  {Tegmark}(2020)}]{Udrescu:2019mnk}%
  \BibitemOpen
  \bibfield  {author} {\bibinfo {author} {\bibfnamefont {S.-M.}\ \bibnamefont
  {Udrescu}}\ and\ \bibinfo {author} {\bibfnamefont {M.}~\bibnamefont
  {Tegmark}},\ }\href {\doibase 10.1126/sciadv.aay2631} {\bibfield  {journal}
  {\bibinfo  {journal} {Sci. Adv.}\ }\textbf {\bibinfo {volume} {6}},\ \bibinfo
  {pages} {eaay2631} (\bibinfo {year} {2020})},\ \Eprint
  {http://arxiv.org/abs/1905.11481} {arXiv:1905.11481 [physics.comp-ph]}
  \BibitemShut {NoStop}%
\bibitem [{\citenamefont {Cranmer}(2023)}]{pysr}%
  \BibitemOpen
  \bibfield  {author} {\bibinfo {author} {\bibfnamefont {M.}~\bibnamefont
  {Cranmer}},\ }\href {https://arxiv.org/abs/2305.01582} {\enquote {\bibinfo
  {title} {Interpretable machine learning for science with pysr and
  symbolicregression.jl},}\ } (\bibinfo {year} {2023}),\ \Eprint
  {http://arxiv.org/abs/2305.01582} {arXiv:2305.01582 [astro-ph.IM]}
  \BibitemShut {NoStop}%
\bibitem [{\citenamefont {Choi}(2011)}]{Choi:2010wa}%
  \BibitemOpen
  \bibfield  {author} {\bibinfo {author} {\bibfnamefont {S.}~\bibnamefont
  {Choi}},\ }\href {\doibase 10.1007/JHEP08(2011)110} {\bibfield  {journal}
  {\bibinfo  {journal} {JHEP}\ }\textbf {\bibinfo {volume} {08}},\ \bibinfo
  {pages} {110} (\bibinfo {year} {2011})},\ \Eprint
  {http://arxiv.org/abs/1006.4998} {arXiv:1006.4998 [hep-ph]} \BibitemShut
  {NoStop}%
\bibitem [{\citenamefont {Butter}\ \emph {et~al.}(2024)\citenamefont {Butter},
  \citenamefont {Plehn}, \citenamefont {Soybelman},\ and\ \citenamefont
  {Brehmer}}]{Butter:2021rvz}%
  \BibitemOpen
  \bibfield  {author} {\bibinfo {author} {\bibfnamefont {A.}~\bibnamefont
  {Butter}}, \bibinfo {author} {\bibfnamefont {T.}~\bibnamefont {Plehn}},
  \bibinfo {author} {\bibfnamefont {N.}~\bibnamefont {Soybelman}}, \ and\
  \bibinfo {author} {\bibfnamefont {J.}~\bibnamefont {Brehmer}},\ }\href
  {\doibase 10.21468/SciPostPhys.16.1.037} {\bibfield  {journal} {\bibinfo
  {journal} {SciPost Phys.}\ }\textbf {\bibinfo {volume} {16}},\ \bibinfo
  {pages} {037} (\bibinfo {year} {2024})},\ \Eprint
  {http://arxiv.org/abs/2109.10414} {arXiv:2109.10414 [hep-ph]} \BibitemShut
  {NoStop}%
\bibitem [{\citenamefont {Dersy}\ \emph {et~al.}(2024)\citenamefont {Dersy},
  \citenamefont {Schwartz},\ and\ \citenamefont {Zhang}}]{Dersy:2022bym}%
  \BibitemOpen
  \bibfield  {author} {\bibinfo {author} {\bibfnamefont {A.}~\bibnamefont
  {Dersy}}, \bibinfo {author} {\bibfnamefont {M.~D.}\ \bibnamefont {Schwartz}},
  \ and\ \bibinfo {author} {\bibfnamefont {X.}~\bibnamefont {Zhang}},\ }\href
  {\doibase 10.1142/S2810939223500028} {\bibfield  {journal} {\bibinfo
  {journal} {Int. J. Data Sci. Math. Sci.}\ }\textbf {\bibinfo {volume} {1}},\
  \bibinfo {pages} {135} (\bibinfo {year} {2024})},\ \Eprint
  {http://arxiv.org/abs/2206.04115} {arXiv:2206.04115 [cs.LG]} \BibitemShut
  {NoStop}%
\bibitem [{\citenamefont {Dong}\ \emph {et~al.}(2023)\citenamefont {Dong},
  \citenamefont {Kong}, \citenamefont {Matchev},\ and\ \citenamefont
  {Matcheva}}]{Dong:2022trn}%
  \BibitemOpen
  \bibfield  {author} {\bibinfo {author} {\bibfnamefont {Z.}~\bibnamefont
  {Dong}}, \bibinfo {author} {\bibfnamefont {K.}~\bibnamefont {Kong}}, \bibinfo
  {author} {\bibfnamefont {K.~T.}\ \bibnamefont {Matchev}}, \ and\ \bibinfo
  {author} {\bibfnamefont {K.}~\bibnamefont {Matcheva}},\ }\href {\doibase
  10.1103/PhysRevD.107.055018} {\bibfield  {journal} {\bibinfo  {journal}
  {Phys. Rev. D}\ }\textbf {\bibinfo {volume} {107}},\ \bibinfo {pages}
  {055018} (\bibinfo {year} {2023})},\ \Eprint
  {http://arxiv.org/abs/2211.08420} {arXiv:2211.08420 [hep-ph]} \BibitemShut
  {NoStop}%
\bibitem [{\citenamefont {Alnuqaydan}\ \emph {et~al.}(2023)\citenamefont
  {Alnuqaydan}, \citenamefont {Gleyzer},\ and\ \citenamefont
  {Prosper}}]{Alnuqaydan:2022ncd}%
  \BibitemOpen
  \bibfield  {author} {\bibinfo {author} {\bibfnamefont {A.}~\bibnamefont
  {Alnuqaydan}}, \bibinfo {author} {\bibfnamefont {S.}~\bibnamefont {Gleyzer}},
  \ and\ \bibinfo {author} {\bibfnamefont {H.}~\bibnamefont {Prosper}},\ }\href
  {\doibase 10.1088/2632-2153/acb2b2} {\bibfield  {journal} {\bibinfo
  {journal} {Mach. Learn. Sci. Tech.}\ }\textbf {\bibinfo {volume} {4}},\
  \bibinfo {pages} {015007} (\bibinfo {year} {2023})},\ \Eprint
  {http://arxiv.org/abs/2206.08901} {arXiv:2206.08901 [hep-ph]} \BibitemShut
  {NoStop}%
\bibitem [{\citenamefont {AbdusSalam}\ \emph {et~al.}(2025)\citenamefont
  {AbdusSalam}, \citenamefont {Abel},\ and\ \citenamefont
  {Crispim~Rom{\~a}o}}]{AbdusSalam:2024obf}%
  \BibitemOpen
  \bibfield  {author} {\bibinfo {author} {\bibfnamefont {S.}~\bibnamefont
  {AbdusSalam}}, \bibinfo {author} {\bibfnamefont {S.}~\bibnamefont {Abel}}, \
  and\ \bibinfo {author} {\bibfnamefont {M.}~\bibnamefont
  {Crispim~Rom{\~a}o}},\ }\href {\doibase 10.1103/PhysRevD.111.015022}
  {\bibfield  {journal} {\bibinfo  {journal} {Phys. Rev. D}\ }\textbf {\bibinfo
  {volume} {111}},\ \bibinfo {pages} {015022} (\bibinfo {year} {2025})},\
  \Eprint {http://arxiv.org/abs/2405.18471} {arXiv:2405.18471 [hep-ph]}
  \BibitemShut {NoStop}%
\bibitem [{\citenamefont {Tsoi}\ \emph {et~al.}(2024)\citenamefont {Tsoi},
  \citenamefont {Pol}, \citenamefont {Loncar}, \citenamefont {Govorkova},
  \citenamefont {Cranmer}, \citenamefont {Dasu}, \citenamefont {Elmer},
  \citenamefont {Harris}, \citenamefont {Ojalvo},\ and\ \citenamefont
  {Pierini}}]{Tsoi:2023isc}%
  \BibitemOpen
  \bibfield  {author} {\bibinfo {author} {\bibfnamefont {H.~F.}\ \bibnamefont
  {Tsoi}}, \bibinfo {author} {\bibfnamefont {A.~A.}\ \bibnamefont {Pol}},
  \bibinfo {author} {\bibfnamefont {V.}~\bibnamefont {Loncar}}, \bibinfo
  {author} {\bibfnamefont {E.}~\bibnamefont {Govorkova}}, \bibinfo {author}
  {\bibfnamefont {M.}~\bibnamefont {Cranmer}}, \bibinfo {author} {\bibfnamefont
  {S.}~\bibnamefont {Dasu}}, \bibinfo {author} {\bibfnamefont {P.}~\bibnamefont
  {Elmer}}, \bibinfo {author} {\bibfnamefont {P.}~\bibnamefont {Harris}},
  \bibinfo {author} {\bibfnamefont {I.}~\bibnamefont {Ojalvo}}, \ and\ \bibinfo
  {author} {\bibfnamefont {M.}~\bibnamefont {Pierini}},\ }\href {\doibase
  10.1051/epjconf/202429509036} {\bibfield  {journal} {\bibinfo  {journal} {EPJ
  Web Conf.}\ }\textbf {\bibinfo {volume} {295}},\ \bibinfo {pages} {09036}
  (\bibinfo {year} {2024})},\ \Eprint {http://arxiv.org/abs/2305.04099}
  {arXiv:2305.04099 [cs.LG]} \BibitemShut {NoStop}%
\bibitem [{\citenamefont {Tsoi}\ \emph
  {et~al.}(2025{\natexlab{a}})\citenamefont {Tsoi}, \citenamefont {Rankin},
  \citenamefont {Caillol}, \citenamefont {Cranmer}, \citenamefont {Dasu},
  \citenamefont {Duarte}, \citenamefont {Harris}, \citenamefont {Lipeles},\
  and\ \citenamefont {Loncar}}]{Tsoi:2024pbn}%
  \BibitemOpen
  \bibfield  {author} {\bibinfo {author} {\bibfnamefont {H.~F.}\ \bibnamefont
  {Tsoi}}, \bibinfo {author} {\bibfnamefont {D.}~\bibnamefont {Rankin}},
  \bibinfo {author} {\bibfnamefont {C.}~\bibnamefont {Caillol}}, \bibinfo
  {author} {\bibfnamefont {M.}~\bibnamefont {Cranmer}}, \bibinfo {author}
  {\bibfnamefont {S.}~\bibnamefont {Dasu}}, \bibinfo {author} {\bibfnamefont
  {J.}~\bibnamefont {Duarte}}, \bibinfo {author} {\bibfnamefont
  {P.}~\bibnamefont {Harris}}, \bibinfo {author} {\bibfnamefont
  {E.}~\bibnamefont {Lipeles}}, \ and\ \bibinfo {author} {\bibfnamefont
  {V.}~\bibnamefont {Loncar}},\ }\href {\doibase 10.1007/s41781-025-00140-9}
  {\bibfield  {journal} {\bibinfo  {journal} {Comput. Softw. Big Sci.}\
  }\textbf {\bibinfo {volume} {9}},\ \bibinfo {pages} {12} (\bibinfo {year}
  {2025}{\natexlab{a}})},\ \Eprint {http://arxiv.org/abs/2411.09851}
  {arXiv:2411.09851 [hep-ex]} \BibitemShut {NoStop}%
\bibitem [{\citenamefont {Soybelman}\ \emph {et~al.}(2024)\citenamefont
  {Soybelman}, \citenamefont {Schiavi}, \citenamefont {Di~Bello},\ and\
  \citenamefont {Gross}}]{Soybelman:2024mbv}%
  \BibitemOpen
  \bibfield  {author} {\bibinfo {author} {\bibfnamefont {N.}~\bibnamefont
  {Soybelman}}, \bibinfo {author} {\bibfnamefont {C.}~\bibnamefont {Schiavi}},
  \bibinfo {author} {\bibfnamefont {F.~A.}\ \bibnamefont {Di~Bello}}, \ and\
  \bibinfo {author} {\bibfnamefont {E.}~\bibnamefont {Gross}},\ }\href
  {\doibase 10.1088/2632-2153/ad8f12} {\bibfield  {journal} {\bibinfo
  {journal} {Mach. Learn. Sci. Tech.}\ }\textbf {\bibinfo {volume} {5}},\
  \bibinfo {pages} {045042} (\bibinfo {year} {2024})},\ \Eprint
  {http://arxiv.org/abs/2406.16752} {arXiv:2406.16752 [hep-ex]} \BibitemShut
  {NoStop}%
\bibitem [{\citenamefont {Dotson}\ \emph {et~al.}(2025)\citenamefont {Dotson}
  \emph {et~al.}}]{Dotson:2025omi}%
  \BibitemOpen
  \bibfield  {author} {\bibinfo {author} {\bibfnamefont {A.}~\bibnamefont
  {Dotson}} \emph {et~al.},\ }\href@noop {} {\  (\bibinfo {year} {2025})},\
  \Eprint {http://arxiv.org/abs/2504.13289} {arXiv:2504.13289 [hep-ph]}
  \BibitemShut {NoStop}%
\bibitem [{\citenamefont {Bahl}\ \emph {et~al.}(2025)\citenamefont {Bahl},
  \citenamefont {Fuchs}, \citenamefont {Menen},\ and\ \citenamefont
  {Plehn}}]{Bahl:2025jtk}%
  \BibitemOpen
  \bibfield  {author} {\bibinfo {author} {\bibfnamefont {H.}~\bibnamefont
  {Bahl}}, \bibinfo {author} {\bibfnamefont {E.}~\bibnamefont {Fuchs}},
  \bibinfo {author} {\bibfnamefont {M.}~\bibnamefont {Menen}}, \ and\ \bibinfo
  {author} {\bibfnamefont {T.}~\bibnamefont {Plehn}},\ }\href@noop {} {\
  (\bibinfo {year} {2025})},\ \Eprint {http://arxiv.org/abs/2507.05858}
  {arXiv:2507.05858 [hep-ph]} \BibitemShut {NoStop}%
\bibitem [{\citenamefont {Vent}\ \emph {et~al.}(2025)\citenamefont {Vent},
  \citenamefont {Winterhalder},\ and\ \citenamefont {Plehn}}]{Vent:2025ddm}%
  \BibitemOpen
  \bibfield  {author} {\bibinfo {author} {\bibfnamefont {S.}~\bibnamefont
  {Vent}}, \bibinfo {author} {\bibfnamefont {R.}~\bibnamefont {Winterhalder}},
  \ and\ \bibinfo {author} {\bibfnamefont {T.}~\bibnamefont {Plehn}},\
  }\href@noop {} {\  (\bibinfo {year} {2025})},\ \Eprint
  {http://arxiv.org/abs/2507.21214} {arXiv:2507.21214 [hep-ph]} \BibitemShut
  {NoStop}%
\bibitem [{\citenamefont {Morales-Alvarado}\ \emph {et~al.}(2024)\citenamefont
  {Morales-Alvarado}, \citenamefont {Conde}, \citenamefont {Bendavid},
  \citenamefont {Sanz},\ and\ \citenamefont
  {Ubiali}}]{Morales-Alvarado:2024jrk}%
  \BibitemOpen
  \bibfield  {author} {\bibinfo {author} {\bibfnamefont {M.}~\bibnamefont
  {Morales-Alvarado}}, \bibinfo {author} {\bibfnamefont {D.}~\bibnamefont
  {Conde}}, \bibinfo {author} {\bibfnamefont {J.}~\bibnamefont {Bendavid}},
  \bibinfo {author} {\bibfnamefont {V.}~\bibnamefont {Sanz}}, \ and\ \bibinfo
  {author} {\bibfnamefont {M.}~\bibnamefont {Ubiali}},\ }in\ \href@noop {}
  {\emph {\bibinfo {booktitle} {{38th conference on Neural Information
  Processing Systems}}}}\ (\bibinfo {year} {2024})\ \Eprint
  {http://arxiv.org/abs/2412.07839} {arXiv:2412.07839 [hep-ph]} \BibitemShut
  {NoStop}%
\bibitem [{\citenamefont {Bendavid}\ \emph {et~al.}(2025)\citenamefont
  {Bendavid}, \citenamefont {Conde}, \citenamefont {Morales-Alvarado},
  \citenamefont {Sanz},\ and\ \citenamefont {Ubiali}}]{Bendavid:2025urn}%
  \BibitemOpen
  \bibfield  {author} {\bibinfo {author} {\bibfnamefont {J.}~\bibnamefont
  {Bendavid}}, \bibinfo {author} {\bibfnamefont {D.}~\bibnamefont {Conde}},
  \bibinfo {author} {\bibfnamefont {M.}~\bibnamefont {Morales-Alvarado}},
  \bibinfo {author} {\bibfnamefont {V.}~\bibnamefont {Sanz}}, \ and\ \bibinfo
  {author} {\bibfnamefont {M.}~\bibnamefont {Ubiali}},\ }\href@noop {} {\
  (\bibinfo {year} {2025})},\ \Eprint {http://arxiv.org/abs/2508.00989}
  {arXiv:2508.00989 [hep-ph]} \BibitemShut {NoStop}%
\bibitem [{\citenamefont {Tsoi}\ \emph
  {et~al.}(2025{\natexlab{b}})\citenamefont {Tsoi}, \citenamefont {Loncar},
  \citenamefont {Dasu},\ and\ \citenamefont {Harris}}]{Tsoi:2024ypg}%
  \BibitemOpen
  \bibfield  {author} {\bibinfo {author} {\bibfnamefont {H.~F.}\ \bibnamefont
  {Tsoi}}, \bibinfo {author} {\bibfnamefont {V.}~\bibnamefont {Loncar}},
  \bibinfo {author} {\bibfnamefont {S.}~\bibnamefont {Dasu}}, \ and\ \bibinfo
  {author} {\bibfnamefont {P.}~\bibnamefont {Harris}},\ }\href {\doibase
  10.1088/2632-2153/adaad8} {\bibfield  {journal} {\bibinfo  {journal} {Mach.
  Learn. Sci. Tech.}\ }\textbf {\bibinfo {volume} {6}},\ \bibinfo {pages}
  {015021} (\bibinfo {year} {2025}{\natexlab{b}})},\ \Eprint
  {http://arxiv.org/abs/2401.09949} {arXiv:2401.09949 [cs.LG]} \BibitemShut
  {NoStop}%
\bibitem [{\citenamefont {Virgolin}\ and\ \citenamefont
  {Pissis}(2022)}]{virgolin2022symbolicregressionnphard}%
  \BibitemOpen
  \bibfield  {author} {\bibinfo {author} {\bibfnamefont {M.}~\bibnamefont
  {Virgolin}}\ and\ \bibinfo {author} {\bibfnamefont {S.~P.}\ \bibnamefont
  {Pissis}},\ }\href {https://arxiv.org/abs/2207.01018} {\enquote {\bibinfo
  {title} {Symbolic regression is np-hard},}\ } (\bibinfo {year} {2022}),\
  \Eprint {http://arxiv.org/abs/2207.01018} {arXiv:2207.01018 [cs.NE]}
  \BibitemShut {NoStop}%
\bibitem [{\citenamefont {Shojaee}\ \emph {et~al.}(2025)\citenamefont
  {Shojaee}, \citenamefont {Meidani}, \citenamefont {Gupta}, \citenamefont
  {Farimani},\ and\ \citenamefont {Reddy}}]{llmsr}%
  \BibitemOpen
  \bibfield  {author} {\bibinfo {author} {\bibfnamefont {P.}~\bibnamefont
  {Shojaee}}, \bibinfo {author} {\bibfnamefont {K.}~\bibnamefont {Meidani}},
  \bibinfo {author} {\bibfnamefont {S.}~\bibnamefont {Gupta}}, \bibinfo
  {author} {\bibfnamefont {A.~B.}\ \bibnamefont {Farimani}}, \ and\ \bibinfo
  {author} {\bibfnamefont {C.~K.}\ \bibnamefont {Reddy}},\ }\href
  {https://arxiv.org/abs/2404.18400} {\enquote {\bibinfo {title} {Llm-sr:
  Scientific equation discovery via programming with large language models},}\
  } (\bibinfo {year} {2025}),\ \Eprint {http://arxiv.org/abs/2404.18400}
  {arXiv:2404.18400 [cs.LG]} \BibitemShut {NoStop}%
\bibitem [{\citenamefont {{Python Software Foundation}}(2023)}]{python}%
  \BibitemOpen
  \bibfield  {author} {\bibinfo {author} {\bibnamefont {{Python Software
  Foundation}}},\ }\href {https://www.python.org/} {\emph {\bibinfo {title}
  {Python: A dynamic, open source programming language}}} (\bibinfo {year}
  {2023})\BibitemShut {NoStop}%
\bibitem [{\citenamefont {Harris}\ \emph {et~al.}(2020)\citenamefont {Harris},
  \citenamefont {Millman}, \citenamefont {van~der Walt}, \citenamefont
  {Gommers}, \citenamefont {Virtanen}, \citenamefont {Cournapeau},
  \citenamefont {Wieser}, \citenamefont {Taylor}, \citenamefont {Berg},
  \citenamefont {Smith}, \citenamefont {Kern}, \citenamefont {Picus},
  \citenamefont {Hoyer}, \citenamefont {van Kerkwijk}, \citenamefont {Brett},
  \citenamefont {Haldane}, \citenamefont {del R{\'{i}}o}, \citenamefont
  {Wiebe}, \citenamefont {Peterson}, \citenamefont {G{\'{e}}rard-Marchant},
  \citenamefont {Sheppard}, \citenamefont {Reddy}, \citenamefont {Weckesser},
  \citenamefont {Abbasi}, \citenamefont {Gohlke},\ and\ \citenamefont
  {Oliphant}}]{numpy}%
  \BibitemOpen
  \bibfield  {author} {\bibinfo {author} {\bibfnamefont {C.~R.}\ \bibnamefont
  {Harris}}, \bibinfo {author} {\bibfnamefont {K.~J.}\ \bibnamefont {Millman}},
  \bibinfo {author} {\bibfnamefont {S.~J.}\ \bibnamefont {van~der Walt}},
  \bibinfo {author} {\bibfnamefont {R.}~\bibnamefont {Gommers}}, \bibinfo
  {author} {\bibfnamefont {P.}~\bibnamefont {Virtanen}}, \bibinfo {author}
  {\bibfnamefont {D.}~\bibnamefont {Cournapeau}}, \bibinfo {author}
  {\bibfnamefont {E.}~\bibnamefont {Wieser}}, \bibinfo {author} {\bibfnamefont
  {J.}~\bibnamefont {Taylor}}, \bibinfo {author} {\bibfnamefont
  {S.}~\bibnamefont {Berg}}, \bibinfo {author} {\bibfnamefont {N.~J.}\
  \bibnamefont {Smith}}, \bibinfo {author} {\bibfnamefont {R.}~\bibnamefont
  {Kern}}, \bibinfo {author} {\bibfnamefont {M.}~\bibnamefont {Picus}},
  \bibinfo {author} {\bibfnamefont {S.}~\bibnamefont {Hoyer}}, \bibinfo
  {author} {\bibfnamefont {M.~H.}\ \bibnamefont {van Kerkwijk}}, \bibinfo
  {author} {\bibfnamefont {M.}~\bibnamefont {Brett}}, \bibinfo {author}
  {\bibfnamefont {A.}~\bibnamefont {Haldane}}, \bibinfo {author} {\bibfnamefont
  {J.~F.}\ \bibnamefont {del R{\'{i}}o}}, \bibinfo {author} {\bibfnamefont
  {M.}~\bibnamefont {Wiebe}}, \bibinfo {author} {\bibfnamefont
  {P.}~\bibnamefont {Peterson}}, \bibinfo {author} {\bibfnamefont
  {P.}~\bibnamefont {G{\'{e}}rard-Marchant}}, \bibinfo {author} {\bibfnamefont
  {K.}~\bibnamefont {Sheppard}}, \bibinfo {author} {\bibfnamefont
  {T.}~\bibnamefont {Reddy}}, \bibinfo {author} {\bibfnamefont
  {W.}~\bibnamefont {Weckesser}}, \bibinfo {author} {\bibfnamefont
  {H.}~\bibnamefont {Abbasi}}, \bibinfo {author} {\bibfnamefont
  {C.}~\bibnamefont {Gohlke}}, \ and\ \bibinfo {author} {\bibfnamefont {T.~E.}\
  \bibnamefont {Oliphant}},\ }\href {\doibase 10.1038/s41586-020-2649-2}
  {\bibfield  {journal} {\bibinfo  {journal} {Nature}\ }\textbf {\bibinfo
  {volume} {585}},\ \bibinfo {pages} {357} (\bibinfo {year}
  {2020})}\BibitemShut {NoStop}%
\bibitem [{\citenamefont {Jones}\ \emph {et~al.}(01  )\citenamefont {Jones},
  \citenamefont {Oliphant}, \citenamefont {Peterson} \emph {et~al.}}]{scipy}%
  \BibitemOpen
  \bibfield  {author} {\bibinfo {author} {\bibfnamefont {E.}~\bibnamefont
  {Jones}}, \bibinfo {author} {\bibfnamefont {T.}~\bibnamefont {Oliphant}},
  \bibinfo {author} {\bibfnamefont {P.}~\bibnamefont {Peterson}},  \emph
  {et~al.},\ }\href {http://www.scipy.org/} {\enquote {\bibinfo {title}
  {{SciPy}: Open source scientific tools for {Python}},}\ } (\bibinfo {year}
  {2001--})\BibitemShut {NoStop}%
\bibitem [{\citenamefont {Paszke}\ \emph {et~al.}(2019)\citenamefont {Paszke},
  \citenamefont {Gross}, \citenamefont {Massa}, \citenamefont {Lerer},
  \citenamefont {Bradbury}, \citenamefont {Chanan}, \citenamefont {Killeen},
  \citenamefont {Lin}, \citenamefont {Gimelshein}, \citenamefont {Antiga},
  \citenamefont {Desmaison}, \citenamefont {Köpf}, \citenamefont {Yang},
  \citenamefont {DeVito}, \citenamefont {Raison}, \citenamefont {Tejani},
  \citenamefont {Chilamkurthy}, \citenamefont {Steiner}, \citenamefont {Fang},
  \citenamefont {Bai},\ and\ \citenamefont {Chintala}}]{pytorch}%
  \BibitemOpen
  \bibfield  {author} {\bibinfo {author} {\bibfnamefont {A.}~\bibnamefont
  {Paszke}}, \bibinfo {author} {\bibfnamefont {S.}~\bibnamefont {Gross}},
  \bibinfo {author} {\bibfnamefont {F.}~\bibnamefont {Massa}}, \bibinfo
  {author} {\bibfnamefont {A.}~\bibnamefont {Lerer}}, \bibinfo {author}
  {\bibfnamefont {J.}~\bibnamefont {Bradbury}}, \bibinfo {author}
  {\bibfnamefont {G.}~\bibnamefont {Chanan}}, \bibinfo {author} {\bibfnamefont
  {T.}~\bibnamefont {Killeen}}, \bibinfo {author} {\bibfnamefont
  {Z.}~\bibnamefont {Lin}}, \bibinfo {author} {\bibfnamefont {N.}~\bibnamefont
  {Gimelshein}}, \bibinfo {author} {\bibfnamefont {L.}~\bibnamefont {Antiga}},
  \bibinfo {author} {\bibfnamefont {A.}~\bibnamefont {Desmaison}}, \bibinfo
  {author} {\bibfnamefont {A.}~\bibnamefont {Köpf}}, \bibinfo {author}
  {\bibfnamefont {E.}~\bibnamefont {Yang}}, \bibinfo {author} {\bibfnamefont
  {Z.}~\bibnamefont {DeVito}}, \bibinfo {author} {\bibfnamefont
  {M.}~\bibnamefont {Raison}}, \bibinfo {author} {\bibfnamefont
  {A.}~\bibnamefont {Tejani}}, \bibinfo {author} {\bibfnamefont
  {S.}~\bibnamefont {Chilamkurthy}}, \bibinfo {author} {\bibfnamefont
  {B.}~\bibnamefont {Steiner}}, \bibinfo {author} {\bibfnamefont
  {L.}~\bibnamefont {Fang}}, \bibinfo {author} {\bibfnamefont {J.}~\bibnamefont
  {Bai}}, \ and\ \bibinfo {author} {\bibfnamefont {S.}~\bibnamefont
  {Chintala}},\ }\href {https://arxiv.org/abs/1912.01703} {\enquote {\bibinfo
  {title} {Pytorch: An imperative style, high-performance deep learning
  library},}\ } (\bibinfo {year} {2019}),\ \Eprint
  {http://arxiv.org/abs/1912.01703} {arXiv:1912.01703 [cs.LG]} \BibitemShut
  {NoStop}%
\bibitem [{\citenamefont {Alwall}\ \emph {et~al.}(2014)\citenamefont {Alwall},
  \citenamefont {Frederix}, \citenamefont {Frixione}, \citenamefont {Hirschi},
  \citenamefont {Maltoni}, \citenamefont {Mattelaer}, \citenamefont {Shao},
  \citenamefont {Stelzer}, \citenamefont {Torrielli},\ and\ \citenamefont
  {Zaro}}]{Alwall:2014hca}%
  \BibitemOpen
  \bibfield  {author} {\bibinfo {author} {\bibfnamefont {J.}~\bibnamefont
  {Alwall}}, \bibinfo {author} {\bibfnamefont {R.}~\bibnamefont {Frederix}},
  \bibinfo {author} {\bibfnamefont {S.}~\bibnamefont {Frixione}}, \bibinfo
  {author} {\bibfnamefont {V.}~\bibnamefont {Hirschi}}, \bibinfo {author}
  {\bibfnamefont {F.}~\bibnamefont {Maltoni}}, \bibinfo {author} {\bibfnamefont
  {O.}~\bibnamefont {Mattelaer}}, \bibinfo {author} {\bibfnamefont {H.~S.}\
  \bibnamefont {Shao}}, \bibinfo {author} {\bibfnamefont {T.}~\bibnamefont
  {Stelzer}}, \bibinfo {author} {\bibfnamefont {P.}~\bibnamefont {Torrielli}},
  \ and\ \bibinfo {author} {\bibfnamefont {M.}~\bibnamefont {Zaro}},\ }\href
  {\doibase 10.1007/JHEP07(2014)079} {\bibfield  {journal} {\bibinfo  {journal}
  {JHEP}\ }\textbf {\bibinfo {volume} {07}},\ \bibinfo {pages} {079} (\bibinfo
  {year} {2014})},\ \Eprint {http://arxiv.org/abs/1405.0301} {arXiv:1405.0301
  [hep-ph]} \BibitemShut {NoStop}%
\bibitem [{\citenamefont {Frederix}\ \emph {et~al.}(2018)\citenamefont
  {Frederix}, \citenamefont {Frixione}, \citenamefont {Hirschi}, \citenamefont
  {Pagani}, \citenamefont {Shao},\ and\ \citenamefont
  {Zaro}}]{Frederix:2018nkq}%
  \BibitemOpen
  \bibfield  {author} {\bibinfo {author} {\bibfnamefont {R.}~\bibnamefont
  {Frederix}}, \bibinfo {author} {\bibfnamefont {S.}~\bibnamefont {Frixione}},
  \bibinfo {author} {\bibfnamefont {V.}~\bibnamefont {Hirschi}}, \bibinfo
  {author} {\bibfnamefont {D.}~\bibnamefont {Pagani}}, \bibinfo {author}
  {\bibfnamefont {H.~S.}\ \bibnamefont {Shao}}, \ and\ \bibinfo {author}
  {\bibfnamefont {M.}~\bibnamefont {Zaro}},\ }\href {\doibase
  10.1007/JHEP11(2021)085} {\bibfield  {journal} {\bibinfo  {journal} {JHEP}\
  }\textbf {\bibinfo {volume} {07}},\ \bibinfo {pages} {185} (\bibinfo {year}
  {2018})},\ \bibinfo {note} {[Erratum: JHEP 11, 085 (2021)]},\ \Eprint
  {http://arxiv.org/abs/1804.10017} {arXiv:1804.10017 [hep-ph]} \BibitemShut
  {NoStop}%
\bibitem [{\citenamefont {Mirkes}(1992)}]{Mirkes:1992hu}%
  \BibitemOpen
  \bibfield  {author} {\bibinfo {author} {\bibfnamefont {E.}~\bibnamefont
  {Mirkes}},\ }\href {\doibase 10.1016/0550-3213(92)90046-E} {\bibfield
  {journal} {\bibinfo  {journal} {Nucl. Phys. B}\ }\textbf {\bibinfo {volume}
  {387}},\ \bibinfo {pages} {3} (\bibinfo {year} {1992})}\BibitemShut {NoStop}%
\bibitem [{\citenamefont {Mirkes}\ and\ \citenamefont
  {Ohnemus}(1995)}]{Mirkes:1994dp}%
  \BibitemOpen
  \bibfield  {author} {\bibinfo {author} {\bibfnamefont {E.}~\bibnamefont
  {Mirkes}}\ and\ \bibinfo {author} {\bibfnamefont {J.}~\bibnamefont
  {Ohnemus}},\ }\href {\doibase 10.1103/PhysRevD.51.4891} {\bibfield  {journal}
  {\bibinfo  {journal} {Phys. Rev. D}\ }\textbf {\bibinfo {volume} {51}},\
  \bibinfo {pages} {4891} (\bibinfo {year} {1995})},\ \Eprint
  {http://arxiv.org/abs/hep-ph/9412289} {arXiv:hep-ph/9412289} \BibitemShut
  {NoStop}%
\bibitem [{\citenamefont {Mirkes}\ and\ \citenamefont
  {Ohnemus}(1994{\natexlab{a}})}]{Mirkes:1994eb}%
  \BibitemOpen
  \bibfield  {author} {\bibinfo {author} {\bibfnamefont {E.}~\bibnamefont
  {Mirkes}}\ and\ \bibinfo {author} {\bibfnamefont {J.}~\bibnamefont
  {Ohnemus}},\ }\href {\doibase 10.1103/PhysRevD.50.5692} {\bibfield  {journal}
  {\bibinfo  {journal} {Phys. Rev. D}\ }\textbf {\bibinfo {volume} {50}},\
  \bibinfo {pages} {5692} (\bibinfo {year} {1994}{\natexlab{a}})},\ \Eprint
  {http://arxiv.org/abs/hep-ph/9406381} {arXiv:hep-ph/9406381} \BibitemShut
  {NoStop}%
\bibitem [{\citenamefont {Mirkes}\ and\ \citenamefont
  {Ohnemus}(1994{\natexlab{b}})}]{Mirkes:1994nr}%
  \BibitemOpen
  \bibfield  {author} {\bibinfo {author} {\bibfnamefont {E.}~\bibnamefont
  {Mirkes}}\ and\ \bibinfo {author} {\bibfnamefont {J.}~\bibnamefont
  {Ohnemus}},\ }in\ \href@noop {} {\emph {\bibinfo {booktitle} {{1994 Meeting
  of the American Physical Society, Division of Particles and Fields (DPF
  94)}}}}\ (\bibinfo {year} {1994})\ pp.\ \bibinfo {pages} {1721--1723},\
  \Eprint {http://arxiv.org/abs/hep-ph/9408402} {arXiv:hep-ph/9408402}
  \BibitemShut {NoStop}%
\bibitem [{\citenamefont {Collins}\ and\ \citenamefont
  {Soper}(1977)}]{Collins:1977iv}%
  \BibitemOpen
  \bibfield  {author} {\bibinfo {author} {\bibfnamefont {J.~C.}\ \bibnamefont
  {Collins}}\ and\ \bibinfo {author} {\bibfnamefont {D.~E.}\ \bibnamefont
  {Soper}},\ }\href {\doibase 10.1103/PhysRevD.16.2219} {\bibfield  {journal}
  {\bibinfo  {journal} {Phys. Rev. D}\ }\textbf {\bibinfo {volume} {16}},\
  \bibinfo {pages} {2219} (\bibinfo {year} {1977})}\BibitemShut {NoStop}%
\bibitem [{\citenamefont {Bern}\ \emph {et~al.}(2011)\citenamefont {Bern} \emph
  {et~al.}}]{Bern:2011ie}%
  \BibitemOpen
  \bibfield  {author} {\bibinfo {author} {\bibfnamefont {Z.}~\bibnamefont
  {Bern}} \emph {et~al.},\ }\href {\doibase 10.1103/PhysRevD.84.034008}
  {\bibfield  {journal} {\bibinfo  {journal} {Phys. Rev. D}\ }\textbf {\bibinfo
  {volume} {84}},\ \bibinfo {pages} {034008} (\bibinfo {year} {2011})},\
  \Eprint {http://arxiv.org/abs/1103.5445} {arXiv:1103.5445 [hep-ph]}
  \BibitemShut {NoStop}%
\bibitem [{\citenamefont {Lam}\ and\ \citenamefont {Tung}(1979)}]{Lam:1978zr}%
  \BibitemOpen
  \bibfield  {author} {\bibinfo {author} {\bibfnamefont {C.~S.}\ \bibnamefont
  {Lam}}\ and\ \bibinfo {author} {\bibfnamefont {W.-K.}\ \bibnamefont {Tung}},\
  }\href {\doibase 10.1016/0370-2693(79)90204-1} {\bibfield  {journal}
  {\bibinfo  {journal} {Phys. Lett. B}\ }\textbf {\bibinfo {volume} {80}},\
  \bibinfo {pages} {228} (\bibinfo {year} {1979})}\BibitemShut {NoStop}%
\bibitem [{\citenamefont {Lam}\ and\ \citenamefont {Tung}(1978)}]{Lam:1978pu}%
  \BibitemOpen
  \bibfield  {author} {\bibinfo {author} {\bibfnamefont {C.~S.}\ \bibnamefont
  {Lam}}\ and\ \bibinfo {author} {\bibfnamefont {W.-K.}\ \bibnamefont {Tung}},\
  }\href {\doibase 10.1103/PhysRevD.18.2447} {\bibfield  {journal} {\bibinfo
  {journal} {Phys. Rev. D}\ }\textbf {\bibinfo {volume} {18}},\ \bibinfo
  {pages} {2447} (\bibinfo {year} {1978})}\BibitemShut {NoStop}%
\bibitem [{\citenamefont {Lam}\ and\ \citenamefont {Tung}(1980)}]{Lam:1980uc}%
  \BibitemOpen
  \bibfield  {author} {\bibinfo {author} {\bibfnamefont {C.~S.}\ \bibnamefont
  {Lam}}\ and\ \bibinfo {author} {\bibfnamefont {W.-K.}\ \bibnamefont {Tung}},\
  }\href {\doibase 10.1103/PhysRevD.21.2712} {\bibfield  {journal} {\bibinfo
  {journal} {Phys. Rev. D}\ }\textbf {\bibinfo {volume} {21}},\ \bibinfo
  {pages} {2712} (\bibinfo {year} {1980})}\BibitemShut {NoStop}%
\bibitem [{\citenamefont {Gehrmann-De~Ridder}\ \emph
  {et~al.}(2016{\natexlab{a}})\citenamefont {Gehrmann-De~Ridder}, \citenamefont
  {Gehrmann}, \citenamefont {Glover}, \citenamefont {Huss},\ and\ \citenamefont
  {Morgan}}]{Gehrmann-DeRidder:2015wbt}%
  \BibitemOpen
  \bibfield  {author} {\bibinfo {author} {\bibfnamefont {A.}~\bibnamefont
  {Gehrmann-De~Ridder}}, \bibinfo {author} {\bibfnamefont {T.}~\bibnamefont
  {Gehrmann}}, \bibinfo {author} {\bibfnamefont {E.~W.~N.}\ \bibnamefont
  {Glover}}, \bibinfo {author} {\bibfnamefont {A.}~\bibnamefont {Huss}}, \ and\
  \bibinfo {author} {\bibfnamefont {T.~A.}\ \bibnamefont {Morgan}},\ }\href
  {\doibase 10.1103/PhysRevLett.117.022001} {\bibfield  {journal} {\bibinfo
  {journal} {Phys. Rev. Lett.}\ }\textbf {\bibinfo {volume} {117}},\ \bibinfo
  {pages} {022001} (\bibinfo {year} {2016}{\natexlab{a}})},\ \Eprint
  {http://arxiv.org/abs/1507.02850} {arXiv:1507.02850 [hep-ph]} \BibitemShut
  {NoStop}%
\bibitem [{\citenamefont {Gehrmann-De~Ridder}\ \emph
  {et~al.}(2016{\natexlab{b}})\citenamefont {Gehrmann-De~Ridder}, \citenamefont
  {Gehrmann}, \citenamefont {Glover}, \citenamefont {Huss},\ and\ \citenamefont
  {Morgan}}]{Gehrmann-DeRidder:2016jns}%
  \BibitemOpen
  \bibfield  {author} {\bibinfo {author} {\bibfnamefont {A.}~\bibnamefont
  {Gehrmann-De~Ridder}}, \bibinfo {author} {\bibfnamefont {T.}~\bibnamefont
  {Gehrmann}}, \bibinfo {author} {\bibfnamefont {E.~W.~N.}\ \bibnamefont
  {Glover}}, \bibinfo {author} {\bibfnamefont {A.}~\bibnamefont {Huss}}, \ and\
  \bibinfo {author} {\bibfnamefont {T.~A.}\ \bibnamefont {Morgan}},\ }\href
  {\doibase 10.1007/JHEP11(2016)094} {\bibfield  {journal} {\bibinfo  {journal}
  {JHEP}\ }\textbf {\bibinfo {volume} {11}},\ \bibinfo {pages} {094} (\bibinfo
  {year} {2016}{\natexlab{b}})},\ \bibinfo {note} {[Erratum: JHEP 10, 126
  (2018)]},\ \Eprint {http://arxiv.org/abs/1610.01843} {arXiv:1610.01843
  [hep-ph]} \BibitemShut {NoStop}%
\bibitem [{\citenamefont {Gauld}\ \emph {et~al.}(2025)\citenamefont {Gauld},
  \citenamefont {Haisch},\ and\ \citenamefont {Weiss}}]{Gauld:2024glt}%
  \BibitemOpen
  \bibfield  {author} {\bibinfo {author} {\bibfnamefont {R.}~\bibnamefont
  {Gauld}}, \bibinfo {author} {\bibfnamefont {U.}~\bibnamefont {Haisch}}, \
  and\ \bibinfo {author} {\bibfnamefont {J.}~\bibnamefont {Weiss}},\ }\href
  {\doibase 10.21468/SciPostPhys.18.5.148} {\bibfield  {journal} {\bibinfo
  {journal} {SciPost Phys.}\ }\textbf {\bibinfo {volume} {18}},\ \bibinfo
  {pages} {148} (\bibinfo {year} {2025})},\ \Eprint
  {http://arxiv.org/abs/2412.13014} {arXiv:2412.13014 [hep-ph]} \BibitemShut
  {NoStop}%
\bibitem [{\citenamefont {Piloneta}\ and\ \citenamefont
  {Vladimirov}(2024)}]{Piloneta:2024aac}%
  \BibitemOpen
  \bibfield  {author} {\bibinfo {author} {\bibfnamefont {S.}~\bibnamefont
  {Piloneta}}\ and\ \bibinfo {author} {\bibfnamefont {A.}~\bibnamefont
  {Vladimirov}},\ }\href {\doibase 10.1007/JHEP12(2024)059} {\bibfield
  {journal} {\bibinfo  {journal} {JHEP}\ }\textbf {\bibinfo {volume} {12}},\
  \bibinfo {pages} {059} (\bibinfo {year} {2024})},\ \Eprint
  {http://arxiv.org/abs/2407.06277} {arXiv:2407.06277 [hep-ph]} \BibitemShut
  {NoStop}%
\bibitem [{\citenamefont {Li}\ \emph {et~al.}(2025{\natexlab{a}})\citenamefont
  {Li}, \citenamefont {Yan},\ and\ \citenamefont {Yuan}}]{Li:2024iyj}%
  \BibitemOpen
  \bibfield  {author} {\bibinfo {author} {\bibfnamefont {X.}~\bibnamefont
  {Li}}, \bibinfo {author} {\bibfnamefont {B.}~\bibnamefont {Yan}}, \ and\
  \bibinfo {author} {\bibfnamefont {C.~P.}\ \bibnamefont {Yuan}},\ }\href
  {\doibase 10.1103/PhysRevD.111.073007} {\bibfield  {journal} {\bibinfo
  {journal} {Phys. Rev. D}\ }\textbf {\bibinfo {volume} {111}},\ \bibinfo
  {pages} {073007} (\bibinfo {year} {2025}{\natexlab{a}})},\ \Eprint
  {http://arxiv.org/abs/2405.04069} {arXiv:2405.04069 [hep-ph]} \BibitemShut
  {NoStop}%
\bibitem [{\citenamefont {Li}\ \emph {et~al.}(2025{\natexlab{b}})\citenamefont
  {Li}, \citenamefont {Li},\ and\ \citenamefont {Yan}}]{Li:2025fom}%
  \BibitemOpen
  \bibfield  {author} {\bibinfo {author} {\bibfnamefont {G.}~\bibnamefont
  {Li}}, \bibinfo {author} {\bibfnamefont {X.}~\bibnamefont {Li}}, \ and\
  \bibinfo {author} {\bibfnamefont {B.}~\bibnamefont {Yan}},\ }\href@noop {} {\
   (\bibinfo {year} {2025}{\natexlab{b}})},\ \Eprint
  {http://arxiv.org/abs/2503.17663} {arXiv:2503.17663 [hep-ph]} \BibitemShut
  {NoStop}%
\bibitem [{\citenamefont {Arroyo-Castro}\ \emph {et~al.}(2025)\citenamefont
  {Arroyo-Castro}, \citenamefont {Scimemi},\ and\ \citenamefont
  {Vladimirov}}]{Arroyo-Castro:2025slx}%
  \BibitemOpen
  \bibfield  {author} {\bibinfo {author} {\bibfnamefont {A.}~\bibnamefont
  {Arroyo-Castro}}, \bibinfo {author} {\bibfnamefont {I.}~\bibnamefont
  {Scimemi}}, \ and\ \bibinfo {author} {\bibfnamefont {A.}~\bibnamefont
  {Vladimirov}},\ }\href {\doibase 10.1007/JHEP06(2025)202} {\bibfield
  {journal} {\bibinfo  {journal} {JHEP}\ }\textbf {\bibinfo {volume} {06}},\
  \bibinfo {pages} {202} (\bibinfo {year} {2025})},\ \Eprint
  {http://arxiv.org/abs/2503.24336} {arXiv:2503.24336 [hep-ph]} \BibitemShut
  {NoStop}%
\end{thebibliography}%
\end{document}